\documentclass{jfm}
\usepackage{natbib}
\usepackage{hyperref}

\usepackage{graphicx}
\usepackage{epstopdf, epsfig}
\usepackage{amsmath}
\usepackage{xcolor}

\shorttitle{Chemotactic cell aggregates as active thin droplets}
\shortauthor{G. L. Celora et al.}

\usepackage{amssymb}
\usepackage{soul}
\usepackage{hyperref}
\usepackage{cleveref}
\usepackage{comment  }

\newcommand{\BLI}[1]{\ensuremath{\bar{#1}}}
\newcommand{\Pe}{\text{Pe}}

\newcommand{\BLII}[1]{\ensuremath{\tilde{#1}}}
\usepackage[colorinlistoftodos]{todonotes}

\usepackage{subcaption}
\title{Chemotaxis of cell aggregates: \\ morphology and dynamics of migrating active droplets}
\author{Giulia L.~Celora\aff{1}
  \corresp{\email{giulia.celora@maths.ox.ac.uk}},
  Benjamin J.~Walker\aff{3}, Mohit P.~Dalwadi\aff{2,3} \corresp{\email{mohit.dalwadi@maths.ox.ac.uk}}, Philip Pearce\aff{3,4} \corresp{\email{philip.pearce@ucl.ac.uk}}}

\affiliation{
\aff{1} Wolfson Centre for Mathematical Biology, Mathematical Institute, University of Oxford, Oxford OX2 6GG, UK
\aff{2} Oxford Centre for Industrial and Applied Mathematics, Mathematical Institute, University of Oxford, Oxford OX2 6GG, UK
\aff{3} Department of Mathematics, University College London, London WC1H 0AY, UK
\aff{4} Institute for the Physics of Living Systems, University College London, London WC1H 0AY, UK
}

\begin{document}

\maketitle

\begin{abstract}
Biological tissues have been observed to display emergent fluid-like properties, owing to physical interactions between cells. However, it remains unclear in general how these fluid-like properties affect tissue structure and function. Here, we are motivated by recent experiments in which cell aggregates were observed to behave as active droplets during collective migration along chemical gradients, or chemotaxis. To understand this process, we develop a minimal model of a growing thin active droplet driven by a self-generated chemical gradient. In broad agreement with the experiments, dynamic simulations reveal that chemotacting droplets exhibit proliferation-driven morphological transitions. To fully characterise these transitions, we perform a multiple scales analysis to show that the droplet dynamics follow a sequence of travelling wave solutions defined by a nonlinear eigenvalue problem parametrised by the slowly increasing droplet volume. Our analysis reveals that morphological transitions can occur continuously or through a discontinuous bifurcation. Further asymptotic analysis of the travelling wave problem reveals that these morphological transitions arise from exponentially small (``beyond-all-orders") asymptotic terms that originate from the rear and front contact lines. Moreover, we show that the nature of the transitions is fully determined by two key dimensionless parameters, which quantify the internal stress balance within the droplet and the strength of the coupling between the droplet migration dynamics and the external chemical field. Overall, our results provide a complete characterisation of the morphodynamics of a class of migrating active thin droplets, with implications in a range of biological systems where cell aggregates exhibit fluid-like behaviour.

\end{abstract}

\begin{keywords} 
active fluids, thin films, travelling wave analysis, matched asymptotic expansions, exponential asymptotics
\end{keywords}

\section{Introduction}

Active fluid models have been successful in explaining dynamics in a range of biological systems, from suspensions of biomolecules (such as actin filaments and proteins) within the intracellular space, to individual cells~\citep{loisy_how_2020,loisy_follow_up,shellard_frictiotaxis_2025}, cell collectives~\citep{ford_pattern_2024,camley_emergent_2016,huycke_patterning_2024}, and tissues~\citep{ibrahimi_stabilization_2025,perez-gonzalez_active_2019,monfared_multiphase-field_2025}. Despite the vastly different physical and biological properties of these systems, they all consist of interacting units that are out of equilibrium, \emph{i.e.}, active. Furthermore, they all possess emergent material properties, such as surface tension and viscosity, that play a key role in determining their spatio-temporal organisation and, consequently, their biological function.

Here, we are primarily motivated by a recent study in which an active fluid model was found to explain collective cell migration along chemical gradients \citep{ford_pattern_2024}, a process known as \emph{chemotaxis}. Specifically, through a combination of experiment and theory, it was shown in \citet{ford_pattern_2024} that when cells migrate as dense, cohesive swarms, they possess emergent material properties and behave as active liquid films. This contrasts with classical theories of chemotaxis based on the foundational work of \cite{keller_traveling_1971}, which have been successful in describing the dynamics of self-generated chemotaxis in a dilute cell population~\citep{tweedy_self-generated_2016,tweedy_seeing_2020,ucar_self-generated_2025,bhattacharjee_chemotactic_2021}. In dense regimes, the assumption of purely self-propelled cell motion breaks down, and fluid-like properties such as surface tension can strongly influence both the migration dynamics and morphology of the cell aggregate \citep{ford_pattern_2024,panigrahi_intermittent_2025}. These findings suggest that the emergent fluid mechanics of cell collectives during long-range migration can have significant consequences for their biological function. 

More generally, self-propelled active droplets have been proposed as an effective framework to study the interplay between morphology and migration of active collectives, from single cells to cell groups. In these systems, migrating droplets are driven by gradients in active stresses. These can result from heterogeneity in the activity or surface tension profile driven by some external chemical field~\citep{ford_pattern_2024,suraj_optimal_active,voss2025bipedalchaoticmotionchemically} or interaction with rigid/free interfaces~\citep{loisy_how_2020,martinezcalvo2024morphodynamicssurfaceattachedactivedrops}. In many biologically relevant settings, the slender geometry of these droplets justifies the use of active thin-film theories -- which can be systematically derived from the full Navier–Stokes equations via lubrication approximations -- to describe their dynamic evolution~\citep{ford_pattern_2024,suraj_optimal_active,loisy_how_2020,voss2025bipedalchaoticmotionchemically}. Passive thin film theories have been used extensively to study fluid flows in a range of technological and industrial applications, including coating, painting, and printing \citep{becker_complex_2003,bonn_wetting_2009,kalliadasis_drop_1994,weinstein2004coating,wilson1993levelling,wilson_evaporation_2023,myers1998thin}. 
However, the rich dynamics of active biological droplets~\citep{ford_pattern_2024,voss2025bipedalchaoticmotionchemically} often depart from that of passive droplets on solid substrates. The mathematical analysis of the fluid dynamics of active biological droplets, therefore, poses novel challenges and requires extending techniques that have been developed for the study of classical passive thin films~\citep{bertozzi_dewetting_2001,castrejon-pita_plethora_2015,craster2009dynamics,dallaston_regular_2021,engelnkemper_morphological_2016,limat_three-dimensional_2004, lunz_dynamics_2018,king_moving_2001,moore_nascent_2021,oliver_contact-line_2015,peschka_signatures_2019,tseluiko_electrified_2008,trinh2014pinned,witelski_nonlinear_2020,xu_variational_2016}. By extending such mathematical analysis to the study of active biological droplets, we hope to guide our understanding of the physics of cell migration; particularly, how biological processes such as proliferation and chemotaxis interact with fluid mechanics to shape the spreading and spatio-temporal dynamics of cells and the collective. 

Here, motivated by active thin-film theories developed recently for cellular aggregates, we formulate and analyse a minimal model of a thin active droplet driven by self-generated gradients in an external chemical field, \emph{i.e.} by chemotaxis. Dynamical simulations reveal the existence of distinct migration regimes, as well as an upper bound on the speed of collective migration. Interestingly, we find that this speed limit is governed by a change in the droplet morphology, which transitions from a compact to an elongated profile. A multiple scales analysis reveals that the nature of such morphological transitions can be investigated by studying a travelling-wave nonlinear eigenvalue problem. By deriving asymptotic approximations of these travelling wave solutions, we determine the physical mechanisms and dominant balances underlying the morphodynamics of self-generated chemotaxis of active droplets and the parameter groupings that control the nature of proliferation-driven morphological transitions. We find that the shape and motion of the droplet near the contact line are fundamental to the macroscopic migration dynamics; in particular, analogously to the study of solitary waves in gravity-driven films~\citep{bertozzi_undercompressive_1999,chang_book_wave,kalliadasis_drop_1994}, capturing this behaviour requires accounting for exponentially small perturbations that arise at the contact lines and matching them across distinct asymptotic regions spanning the solution domain.
\section{Problem set-up}
\begin{figure}
    \centering
    \includegraphics[width=0.65\linewidth]{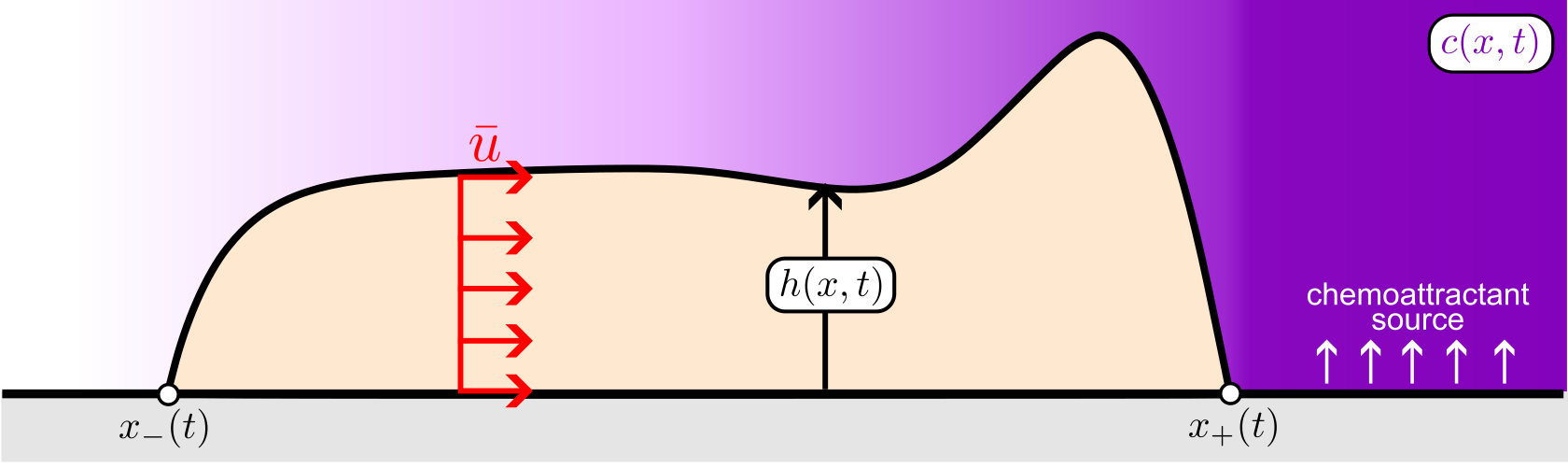}
    \caption{Schematic of the thin active droplet model of collective self-generated chemotaxis of cell groups. The cell group is defined by the evolution of the free-surface $y=h(x,t)$, which touches the substrate at the contact lines $x=x_\pm(t)$. The migration of the active droplet is driven by active pressure gradients coupled to an external chemotactic field, $c=c(x,t)$, which is shaped by the advancing droplet itself via depletion of the chemoattractant source. Since we consider a thin droplet, the chemoattractant is taken to be homogeneous in the vertical direction.}
    \label{fig:model schematic}
\end{figure}

We investigate a minimal continuum model of chemotaxis in dense cell aggregates, formulated as a modified version of the classical thin-film equation for a gravity-driven sliding fluid. The model is adapted from~\citet{ford_pattern_2024}, and emerges from a coarse-grained description of a thin, dense aggregate of cells migrating under the guidance of an external chemical field $c=c(x,t)$, which is self-generated because cells break down the chemical (\Cref{fig:model schematic}). As discussed in~\citet{ford_pattern_2024}, surface tension emerges from unbiased cell-cell interactions and confines cells within the aggregate, and the alignment of directed cell motion owing to chemotactic bias induces an effective active pressure. Gradients of this active pressure across the aggregate act as a forcing that propels the aggregate forward. Within this framework, long-range migration of cell aggregates is therefore achieved by the coupling of active pressure gradients to the external chemoattractant field $c$, whose spatial distribution is shaped by the cells through chemical degradation.

As illustrated in~\Cref{fig:model schematic}, we consider a 2D cross-section of the cell aggregate and describe the boundary of the aggregate by the height function $h=h(x,t)$, which is defined on the moving domain $x\in[x_-(t),x_+(t)]$, where $x_\pm$ denote the contact lines between the aggregate and the substrate. As the droplet advances, we assume its size increases at a constant rate $r$ due to proliferation. Under these assumptions and following the non-dimensionalisation procedure outlined in~\Cref{app:nondim}, the morphodynamics of the chemotactic aggregate is described by the following dimensionless nonlinear fourth-order moving boundary problem for its height profile $h$:
\begin{subequations}
    \begin{align}
        \partial_t h +\partial_x\left(h\bar{u}\right)= rh, \quad x\in[x_-(t),x_+(t)],\ &t>0,\label{eq:time dependent H}
    \end{align}
    where $\bar{u}$ represents the depth-averaged speed of the cells within the aggregate. Assuming perfect slip between the aggregate and the substrate (\emph{i.e.}, allowing cells to move freely on the substrate with negligible friction), $\bar{u}$ takes the form
    \begin{align}\label{eq:average speed}
        \bar{u}=\frac{h}{Ca_\kappa} \left(h_{xxx}+f_a(c(x,t))\right),
    \end{align}\label{general governing equations}%
    \end{subequations}
    where $Ca_\kappa=\bar{U}\eta/\kappa$ is the droplet capillary number representing the ratio between viscous ($\bar{U}\eta $) and surface tension ($\kappa$) effects; here, $\eta$ is the emergent dynamic viscosity of the aggregate, $\kappa$ is its surface tension and $\bar{U}$ is the characteristic migration speed of the droplet. In \eqref{eq:average speed}, the $h_{xxx}$ term captures capillary-driven flows that are controlled by a constant surface tension (here normalised to unity), while the $f_a$ term describes activity-driven flows that, in our model, arise from the heterogeneity in cellular activity owing to the spatially-heterogeneous distribution of the chemoattractant field across the droplet (see schematic in~\Cref{fig:model schematic}). 

    Generally, the $f_a$ term would lead to spatio-temporally heterogeneous active forcing, which is regulated by the migration of the cell aggregate itself because the cells in the aggregate degrade the chemoattractant~\citep{ford_pattern_2024}. However, solving for the chemoattractant evolution~\eqref{sys:chemoattractant} under a quasi-steady approximation, and under the simplifying assumptions of cell chemotaxis being driven by perfect logarithmic sensing (see~\eqref{eq:activityI}), as in the standard Keller-Segel model~\citep{keller_traveling_1971}, we find that the resulting self-generated active force is spatially homogeneous (see~\Cref{app:derivation chemoattractant} for derivation). Importantly, however, its magnitude 
    \begin{align}
    f_a=\frac{\mathcal{A}}{\ell_c(\dot{x}_+\eta_c)}, \label{eq:active forcing chemotaxis}
    \end{align}
    is modulated by the effective decay length of the chemoattractant 
    \begin{align}
        \ell_c(\Pe_c)=\frac{\Pe_c}{2}+\sqrt{\left(\frac{\Pe_c}{2}\right)^2+1},\label{eq:decay chemoattractant}
    \end{align}
    resulting in a nonlinear feedback between the migration speed of the aggregate (via $\dot{x}_+$) and the driving force $f_a$. In Eq.~(\ref{eq:active forcing chemotaxis}), $\mathcal{A}>0$ is a positive constant indicating the maximum value of the function $f_a$ attained when the coupling to the external chemical field is neglected (\emph{i.e.}, $\Pe_c=0$), and measures the ratio between active and capillary forces (see Eq.~\eqref{app: def A} in~\Cref{app:derivation chemoattractant}). The chemoattractant P\'{e}clet number $\Pe_c=\dot{x}_+\eta_c$ is defined as the ratio between the droplet speed $\dot{x}_+$ and the characteristic reaction-diffusion speed of the chemoattractant $\eta^{-1}_c$ (see Eq.~\eqref{app:def etac} in~\Cref{app:derivation chemoattractant}) and measures the extent to which the chemoattractant penetration into the cell aggregate is limited by migration dynamics of the aggregate ($\Pe_c\gg1$) or diffusion ($\Pe_c\ll1$).

    We couple the governing equation~\eqref{eq:time dependent H} with appropriate boundary conditions.
    At the contact lines, the height of the droplet is fixed
    \begin{subequations}
        \begin{equation}
        h(x_\pm(t),t)=0, \quad t\geq0,\label{eq:cond_diric}
    \end{equation}
    and the aggregate forms well-defined contact angles with the substrate, which are regulated dynamically according to:
    \begin{equation}
        \eta_\theta \dot{x}_\pm =  \pm\left[\left(\left.\partial_x h\right|_{x=x_\pm}\right)^2 -1\right],\quad t\geq0.\label{eq:dynamics_contact angle}
    \end{equation}
    In Eq.~(\ref{eq:dynamics_contact angle}), the positive constant $\eta_\theta$ is related to the rate at which energy is dissipated at the contact lines as the droplet advances~\citep{peschka_variational_2018}. Without loss of generality, we here assume that the equilibrium contact angle $\theta_e$ of the droplet satisfies $\tan\theta_e=1$, which corresponds to rescaling the spatial dimension and the height of the droplet by a relative factor $H/L=\tan\theta_e$ (see~\Cref{app:nondim}). Finally, to close the system, we assume that no mass is generated or depleted at the contact line by imposing no-flux conditions
    \begin{align}
        \left.h(\bar{u}-\dot{x}_\pm)\right|_{x=x_\pm}=0, \quad t\geq0.\label{eq:no-flux}
    \end{align}\label{general BC}%
Effectively, assuming that $\dot{x}_\pm$ and $f_a$ are finite, Eqs.~(\ref{eq:cond_diric}) and (\ref{eq:no-flux}) impose that the quantity $h^2h_{xxx}$ vanishes at the contact lines. 
\end{subequations}

Our model is now fully specified and consists of Eqs.~\eqref{general governing equations}-\eqref{eq:decay chemoattractant} and boundary conditions~\eqref{general BC}. The model has a total of 5 non-dimensional parameters (see~\Cref{tab:parameters}): the aggregate growth rate $r$, the passive capillary number $Ca_\kappa$, the active capillary number $\mathcal{A}$, the contact line dissipation parameter $\eta_\theta$, and the chemoattractant coupling parameter $\eta_c$. If $f_a$ were constant (setting $\eta_c=0$) and proliferation neglected by setting $r=0$, Eqs.~(\ref{general governing equations})-\eqref{general BC} reduce to the description of a passive thin droplet sliding under a constant gravitational forcing with dissipation at the contact line, which has been previously studied in~\citet{peschka_variational_2018}, primarily via numerical simulations. Our formulation and subsequent analysis extend this investigation to study the morphodynamics of chemotactic active droplets. 

\begin{figure}
    \centering
       \begin{subfigure}{0.006\textwidth}
    \captionlistentry{}
    \label{fig:SG_dynamicsA_homogeneous}
    \end{subfigure}
    \begin{subfigure}{0.95\textwidth}
    \centering
    \includegraphics[width=\textwidth]{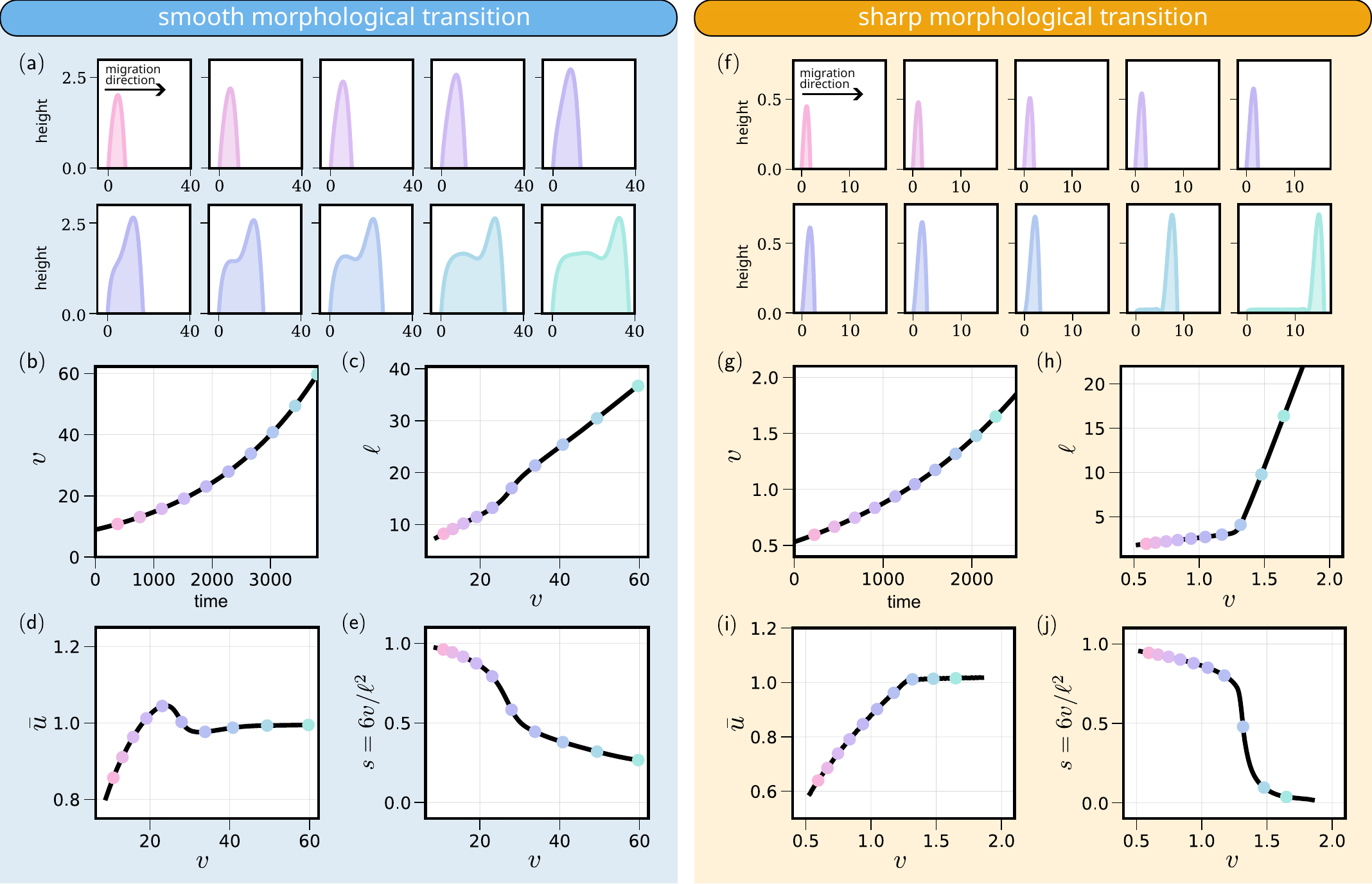}
    \captionlistentry{}
    \label{fig:SG_dynamicsB_homogeneous}
    \end{subfigure}
    \begin{subfigure}{0.001\textwidth}
    \captionlistentry{}
    \label{fig:SG_dynamicsC_homogeneous}
    \end{subfigure}
    \begin{subfigure}{0.001\textwidth}
    \captionlistentry{}
    \label{fig:SG_dynamicsD_homogeneous}
    \end{subfigure}
        \begin{subfigure}{0.001\textwidth}
    \captionlistentry{}
    \label{fig:SG_dynamicsE_homogeneous}
    \end{subfigure}
            \begin{subfigure}{0.001\textwidth}
    \captionlistentry{}
    \label{fig:SG_dynamicsF_homogeneous}
    \end{subfigure}
            \begin{subfigure}{0.001\textwidth}
    \captionlistentry{}
    \label{fig:SG_dynamicsG_homogeneous}
    \end{subfigure}
            \begin{subfigure}{0.001\textwidth}
    \captionlistentry{}
    \label{fig:SG_dynamicsH_homogeneous}
    \end{subfigure}
    \begin{subfigure}{0.001\textwidth}
    \captionlistentry{}
    \label{fig:SG_dynamicsI_homogeneous}
    \end{subfigure}
    \begin{subfigure}{0.001\textwidth}
    \captionlistentry{}
    \label{fig:SG_dynamicsJ_homogeneous}
    \end{subfigure}
    \vspace{-4mm}
    \caption{Representative simulations of morphological transitions in a chemotactic growing droplet~\eqref{general governing equations} (see~\Cref{app:numerics} for details on the numerical implementation). The shaded colors highlight results from two distinct parameter sets corresponding to (blue) a smooth (set 1 in~\Cref{tab:parameters}) and (yellow) a sharp (set 2 in~\Cref{tab:parameters}) morphological transition. (a) and (f) Example of the droplet height profile $h(x,t)$ at distinct times. Time increases from left to right (see corresponding colored dots in other panels). Note that the solution is shifted so that $x=0$ corresponds to the location of the left boundary $x_-(t)$. (b) and (g) Time evolution of the droplet volume, $v=\int_{x_-}^{x_+} \! h(x,t)\, \mathrm{d}x$. (c) and (h) Plots of the droplet length, $\ell=x_+(t)-x_-(t)$ vs droplet volume $v$.  (d) and (i) Plots of the speed of the droplet centre of mass, $u_{\bar{x}}$ vs droplet volume $v$. (e) and (j) Plots of the shape parameter $s$ vs droplet volume $v$. }
    \label{fig:SG_travelling_homogeneous_example}
\end{figure}
\subsection{Numerical simulations}

In Figure~\ref{fig:SG_travelling_homogeneous_example}, we show two representative simulations of the model~\eqref{general governing equations}-\eqref{general BC} corresponding to different parameter choices (see~\Cref{tab:parameters}). Independent of the choice of model parameters, we find that migrating droplets transition from a compact to an elongated morphology as time progresses (Figures~\ref{fig:SG_dynamicsA_homogeneous} and~\ref{fig:SG_dynamicsF_homogeneous}), and the droplet volume increases (Figures~\ref{fig:SG_dynamicsB_homogeneous} and~\ref{fig:SG_dynamicsG_homogeneous}). At early times, when the droplet volume is sufficiently small, the droplet migrates relatively quickly while maintaining a compact shape, with its centre of mass only slightly biased towards the forward direction. As the droplet volume increases, its morphology eventually changes into an elongated and highly asymmetric profile, characterised by a long flat region that connects the rear and front of the droplet, while the peak height approaches a constant value. 

We quantify changes in droplet morphologies by looking at the relationship between droplet volume and three key metrics: the length of the droplet $\ell$ (Figures \ref{fig:SG_dynamicsC_homogeneous} and~\ref{fig:SG_dynamicsH_homogeneous}), the speed of its centre of mass (\Cref{fig:SG_dynamicsD_homogeneous} and~\ref{fig:SG_dynamicsI_homogeneous}) and the shape parameter $s:=6v/\ell^2$ (Figure~\ref{fig:SG_dynamicsE_homogeneous} and~\ref{fig:SG_dynamicsJ_homogeneous}), which quantifies the deviation of the droplet shape from its equilibrium configuration (corresponding to $s=1$) and decreases as the droplet becomes more elongated. These metrics show that the response of the droplet near the critical volume associated with the compact-to-elongated transition are significantly different between the two parameter regimes. In the first scenario (blue shaded region in~\Cref{fig:SG_travelling_homogeneous_example}), changes in droplet morphology as its volume $v$ increases are smooth, as indicated by the gradual decrease in the shape parameter (\Cref{fig:SG_dynamicsE_homogeneous}). When looking at the migration dynamics, we find that the morphological transition to an elongated morphology drives a significant reduction in the migration speed (\Cref{fig:SG_dynamicsD_homogeneous}). Growth beyond a critical volume is therefore disadvantageous for the collective, suggesting that mechanisms for mass control may be advantageous to maintain a maximal migration speed. By contrast, in the second physical regime (yellow shaded region in~\Cref{fig:SG_travelling_homogeneous_example}), changes in the droplet morphology are less gradual, as highlighted by the sharp decrease in the shape parameter $s$ beyond a critical value of the droplet volume (Figure~\ref{fig:SG_dynamicsJ_homogeneous}). Furthermore, the morphological transition to an elongated morphology has a negligible impact on the migration speed (\Cref{fig:SG_dynamicsH_homogeneous}), and instead results in a steeper increase in the length of the droplet with respect to its volume (\Cref{fig:SG_dynamicsG_homogeneous}).

These numerical simulations illustrate the existence of proliferation-driven morphological transitions in our minimal model of self-generated chemotaxis of growing active droplets, broadly in line with those observed experimentally~\citep{ford_pattern_2024}. In our model, these transitions impose an upper limit on the migration speed of the group, defining an optimal droplet size that maximises collective motility. Yet, it remains unclear how the physical and chemical processes in the model regulate the migration speed of the collective and the droplet morphology. With this in mind, our goal is to understand and comprehensively characterise the key features of the migration dynamics; namely, how changes in the droplet volume modulate the migration speed and the droplet morphology by altering the balance between the different physical mechanisms in the model. 

\section{Travelling wave analysis}
We start by systematically reducing Eqs.~(\ref{general governing equations})-\eqref{general BC} to a travelling wave (TW) nonlinear eigenvalue problem for the travelling speed $u$ and droplet morphology $h$, where the droplet volume becomes an effective control parameter. We show that the TW analysis can explain the behaviour observed in the dynamical simulations and further characterise the nature of the transitions driving morphological changes in the profile of migrating droplets. 

\subsection{Derivation of the travelling wave problem}
\label{sec:main multiple scales}
To proceed, we first leverage the separation between the slower characteristic time-scale of cell proliferation, $t_{p}=r^{-1}\gg 1$, and the faster characteristic time-scale of migration, $t_u=\bar{u}^{-1}\sim\mathcal{O}(1)$ as $r\to0$ to formally simplify Eqs.~(\ref{general governing equations})-\eqref{general BC} using the method of multiple scales. As outlined in Appendix~\ref{app_sec:multiple_scales}, in the limit $r\rightarrow 0$, the fast dynamics of the system occur over $t=\mathcal{O}(1)$ wherein the leading-order problem conserves mass, and is determined by:
\begin{subequations}\label{sys:h_multiple_scales}
\begin{align}
    \partial_t h +\partial_x\left(h^2Ca^{-1}_\kappa\left( h_{xxx}+f_a(\partial_t x_+)\right)\right)=0, \quad x\in\left[x_{-}(t;T),x_+(t;T)\right],\\[5pt]
    h^2h_{xxx}=0,\quad h=0,\quad 
    \eta_\theta\partial_t{x}_{\pm}=\pm\left[(\partial_x h)^2-1\right], \quad x=x_\pm(t;T).\label{eq:h_multiple_scales_BC_conditions}
   %   h(x_{\pm}(t),t)=0,\quad    \eta_\theta \dot{x}_\pm- \left[\left(\left.\partial_x h\right|_{x=x_\pm}\right)^2-\tan^2\theta_e\right]n_\pm=0, \quad &t>0,
\end{align}
\end{subequations}
 We note that the leading-order problem~\eqref{sys:h_multiple_scales} has a quasi-steady dependence on the slow time-scale $T=rt$, which only enters via the slowly varying volume constraint 
\begin{equation}\label{eq:volume constaint_slow_variable}
v(T)=\int_{x_-(t;T)}^{x_{+}(t;T)}h(x,t;T)dx=v(0)\exp{(T)}. 
\end{equation}
For the purpose of the following analysis, it is simpler to parameterise the leading-order problem by $v$ rather than $T$; however, given the above expression, the two are interchangeable. 

Over the fast time-scale $t$, our numerical simulations suggest that the solutions relax to a travelling wave solution. To investigate this, we look for travelling wave solutions in the form
\begin{equation}\label{eq:TW ansatz}
h=\tilde{h}(x-u(v)t;v), \quad x_\pm(t;v)=\pm\frac{\ell(v)}{2}+u(v)t,
\end{equation}
where $u(v)$ is the as-yet-unknown travelling wave speed. 
Using the travelling wave ansatz~\eqref{eq:TW ansatz} in~\eqref{sys:h_multiple_scales}, we obtain the following nonlinear eigenvalue problem
\begin{subequations}
\begin{align}
        \tilde{h}'''(z)&=\left[\frac{Ca_\kappa u}{ \tilde{h}(z)} - f_{a}(u)\right],\quad z\in\left(-\frac{\ell}{2},\frac{\ell}{2}\right),\label{eq:H_TW1}\\
        \tilde{h}\left(\pm\frac{\ell}{2}\right)&=0, \quad
         \tilde{h}'\left(\pm\frac{\ell}{2}\right)\pm\sqrt{1\pm \eta_\theta \,u}=0,\\  %H'(1/2)&= -\sqrt{1+Ca_\mathcal{N}_\theta\, \mathcal{U}},\\
       v&=\int_{-\ell/2}^{\ell/2} \tilde{h}(z)\, dz,\label{eq_TW:mass conservation}
    \end{align}\label{eq:general_TW_problem}% 
\end{subequations}
where the travelling speed $u$ is the eigenvalue and, for simplicity of notation, we have dropped the explicit dependence of the variables on $v$. Note that Eq.~(\ref{eq:H_TW1}) automatically satisfies the no-flux condition $h^2h_{xxx}=0$ from the original problem~\eqref{eq:h_multiple_scales_BC_conditions}. As detailed in~\Cref{app:mass calculation}, we can rewrite~\eqref{eq_TW:mass conservation} as an explicit expression linking the volume $v$, speed $u$ and length $\ell$ of the droplet
\begin{equation}
v=\frac{Ca_\kappa u(\ell+ Ca^{-1}_\kappa \eta_\theta)}{f_{a}(u)}.\label{def volume dimensional}
\end{equation}
\subsection{Numerical bifurcation analysis of TW solutions}
\label{sec:numerical continuation}
\begin{figure}
    \centering
     \begin{subfigure}{0.002\textwidth}
    \captionlistentry{}
    \label{fig bifurcation constant forcing panel A}
    \end{subfigure}
        \begin{subfigure}{0.98\textwidth}
    \captionlistentry{}
    \includegraphics[width=\linewidth]{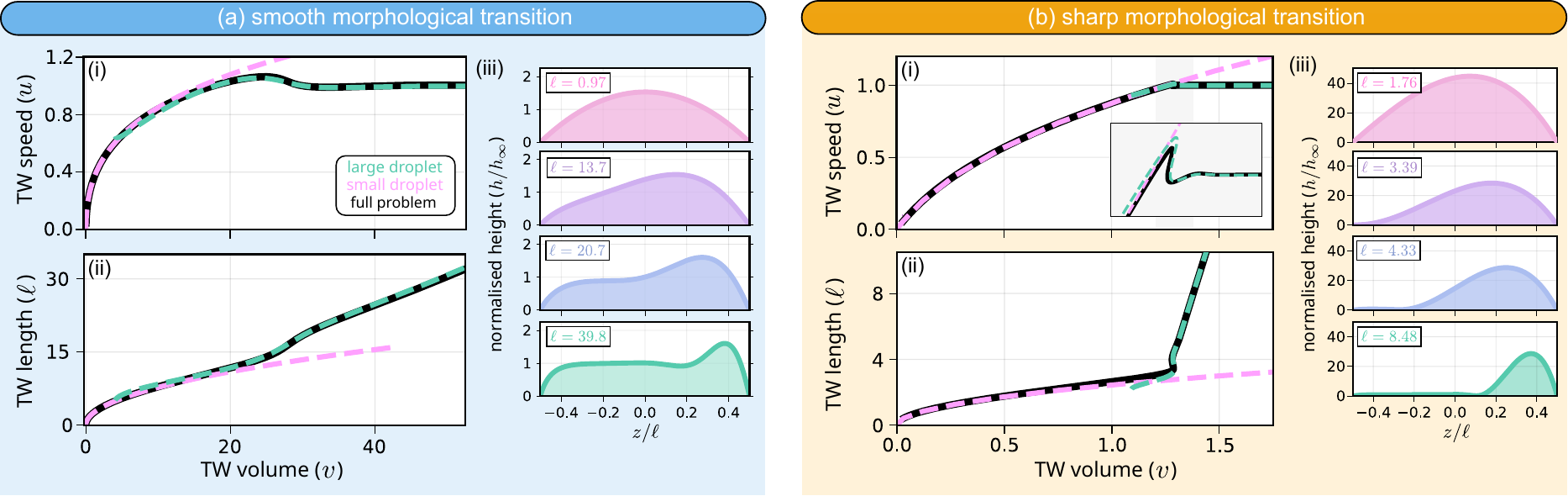}
        \label{fig bifurcation constant forcing panel B}
    \end{subfigure}
    \begin{subfigure}{0.002\textwidth}
    \captionlistentry{}
    \label{fig bifurcation constant forcing panel C}
    \end{subfigure}
    \vspace{-8mm}
    \caption{(a)-(b) Numerical solutions of the full travelling wave problem~\eqref{eq:general_TW_problem} corresponding to the parameter values used in~\Cref{fig:SG_travelling_homogeneous_example}.
    (i)-(ii) Numerical bifurcation diagram (black curve) showing the estimated (i) volume-speed and (ii) volume-length relations. Results are compared with the prediction of the asymptotic approximations (dashed lines) discussed in~\Cref{sec analysis:summary}. (iii) Examples of droplet profile of travelling wave solutions for increasing droplet length ($\ell$).}
    \label{fig bifurcation constant forcing}
\end{figure}

We solve Eq.~(\ref{eq:general_TW_problem}) numerically using continuation techniques (see~\Cref{app:numerical continuation} for details), for the same choice of the parameters as in~\Cref{fig:SG_travelling_homogeneous_example}. Results are illustrated in Figures~\ref{fig bifurcation constant forcing panel A}-\ref{fig bifurcation constant forcing panel B}. The obtained relationship between the travelling speed $u$ (panel (i)), the droplet length $\ell$ (panel (ii)), and volume $v$ agrees with the corresponding behaviour observed in the dynamical simulations (\Cref{fig:SG_travelling_homogeneous_example}). The travelling wave analysis also recapitulates the changes in the morphology of the droplet from a compact to an elongated morphology (panel (iii)). We therefore conclude that the full problem dynamics can be interpreted as the system slowly evolves along the family of travelling wave solutions parameterised by the droplet mass. 

The travelling wave analysis further reveals how parameters influence the nature of the morphological transitions. In the regime where proliferation drives smooth dynamics of the droplet shape, the morphological transition of travelling wave solutions from a compact to an elongated morphology occurs continuously (\Cref{fig bifurcation constant forcing panel A}). By contrast, sharp morphological transitions in the dynamic simulations are mediated by a discontinuous (fold) bifurcation, leading to a range of droplet masses for which multiple travelling wave solutions coexist (\Cref{fig bifurcation constant forcing panel B}). 

We next turn to characterise travelling wave solutions analytically with the aim of better understanding the physics of morphological transitions in migrating droplets and the mechanisms that select for their maximal migration speed. 

\section{Asymptotic analysis of travelling wave solutions}
\label{sec analysis:summary}
The biologically relevant control parameter driving morphological change in migrating droplets is the droplet volume $v$, but the droplet length $\ell$ is a more natural choice of control parameter for studying solutions to the system~\eqref{eq:general_TW_problem} analytically. Once a solution is obtained,~\eqref{def volume dimensional} provides a straightforward relationship between $v$ and $\ell$.
Classical studies of travelling wave problems similar to~\eqref{eq:general_TW_problem} typically investigate the behaviour of small droplet solutions in the small-$\ell$ limit~\citep{peschka_variational_2018,loisy_how_2020}. We conduct an analysis of~\eqref{eq:general_TW_problem} in this small-$\ell$ limit in~\Cref{app: small droplet limit}, using a regular perturbative expansion to compute all unknown quantities up to the first-order approximation.
Although this limit can capture the initial increase in the droplet speed as a function of volume (magenta dashed curves in~\Cref{fig bifurcation constant forcing panel A,fig bifurcation constant forcing panel B}), it fails to capture the saturation in the migration speed and the associated transition to an elongated morphology, which we are interested in studying. By contrast, we find that the appropriate limit to capture the observed morphological transitions is the (singular) large-$\ell$ limit, which will be the focus of this section.

Specifically, we will show that the non-monotonic relationship between droplet volume and, for example, its migration speed (green dashed curves in~\Cref{fig bifurcation constant forcing panel A,fig bifurcation constant forcing panel B}) requires calculating the first-correction term. Importantly, in contrast with the small droplet regime $\ell\rightarrow 0$, the first correction in the large-$\ell$ limit consists of exponentially small terms. These effects technically arise ``beyond-all-orders" of a standard perturbative expansion, though we note that a standard expansion here truncates after finitely many terms.

\subsection{Derivation of elongated droplet solutions: general set-up.}
\label{sec logathmic asymptotics}

\begin{figure}
\begin{subfigure}{0.002\textwidth}
     \centering
        \captionlistentry{}
\label{fig:asymptotic solution structure}
\end{subfigure}
\begin{subfigure}{0.002\textwidth}
     \centering
        \captionlistentry{}
\label{fig:asymptotics work flow}
\end{subfigure}
\begin{subfigure}{0.99\textwidth}
     \centering
    \includegraphics[width=\linewidth]{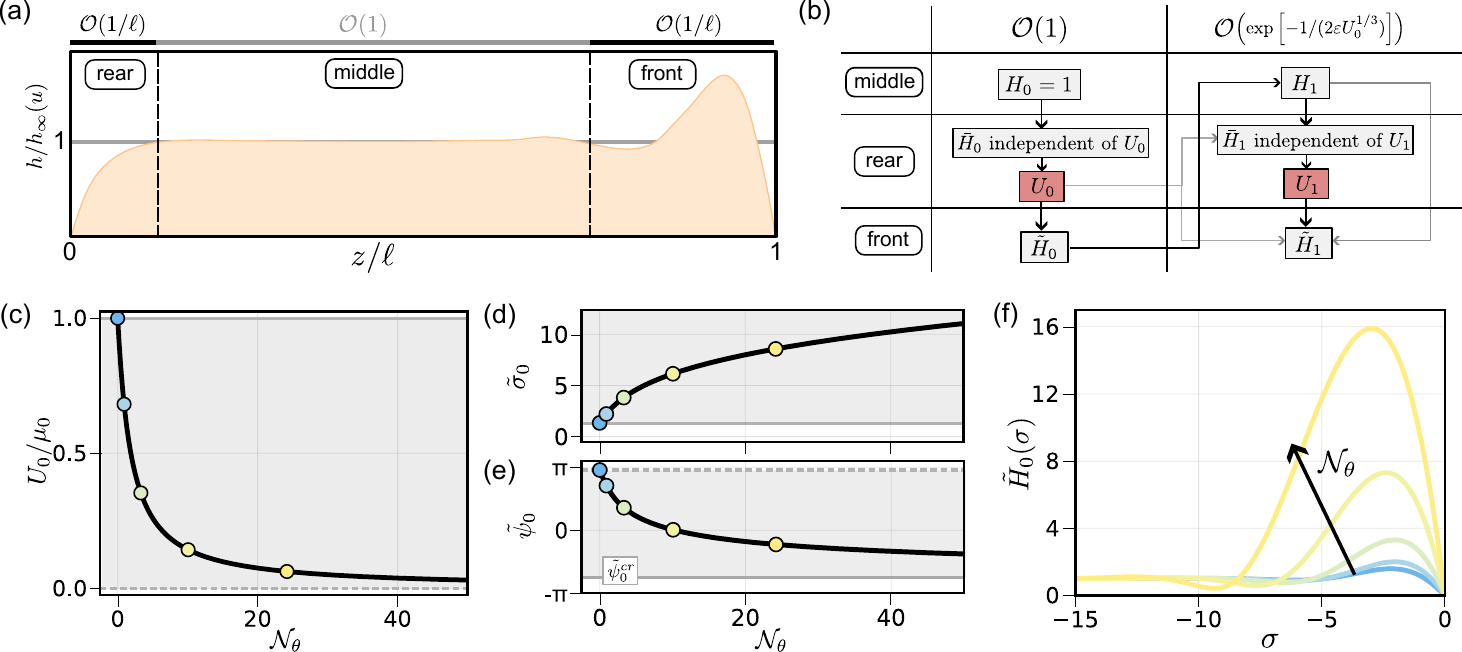}
       \captionlistentry{}
    \label{fig:regionIA}
\end{subfigure}
 \begin{subfigure}{0.002\textwidth}
     \centering
        \captionlistentry{}
    \label{fig:regionIB}
    \end{subfigure}
        \begin{subfigure}{0.002\textwidth}
     \centering
        \captionlistentry{}
    \label{fig:regionIC}
    \end{subfigure}
     \begin{subfigure}{0.002\textwidth}
     \centering
        \captionlistentry{}
    \label{fig:regionID}
    \end{subfigure}
    \vspace{-5mm}
\vspace{-5mm}
    \caption{(a) Schematic of the structure of the droplet morphology in the large droplet limit. Two boundary layers form near the rear and front contact lines, in which the droplet profile rapidly changes until settling to a uniform solution in the middle region connecting the two. (b) A flow chart displaying the information flow for the large droplet asymptotic limit. The black arrows show the direction of the information flow, and the grey arrows indicate secondary connections. The grey and red boxes indicate results associated with the approximation of the travelling droplet profile and speed, respectively. (c)-(f) Characterisation of the leading-order behaviour as a function of the non-dimensional parameter $\mathcal{N}_\theta$ (see~\eqref{eq:def_ntheta}). Mapping between $\mathcal{N}_\theta$ and (c) the leading-order velocity $U_0$ (see~\eqref{eq constant: U_0}), and  (d) $\BLII{\sigma}_0$ and (e) $\BLII{\psi}_0$, which characterise the far-field behaviour of the solution in the front boundary layer (see~\eqref{eq:farfield front}). The grey shaded area indicates the range of values taken by each variable, with dotted yellow curves indicating asymptotes. (f) Examples of drop profiles in the front boundary layer for increasing values of $\mathcal{N}_\theta$.}
    \label{fig:schematic asympotics}
\end{figure}

 Before presenting the details of the analysis, we briefly summarise the asymptotic structure of the solution that we will subsequently identify and analyse. As suggested by the numerically computed droplet profile, and shown schematically in~\Cref{fig:schematic asympotics}, the solution domain divides into three asymptotic regions: the two inner regions near the rear and front contact lines have width $\mathcal{O}(1)$, independent of the length $\ell$ of the droplet, and the outer region that connects the two inner regions has width that scales with $\ell$. The height of the droplet in the outer region corresponds to the equilibrium point, or constant solution of the ODE~(\ref{eq:H_TW1}), namely $h\sim Ca_\kappa u/f_a(u)$. Importantly, we do not know $u$ \emph{a priori} -- we determine $u$ through our subsequent analysis in what is effectively a nonlinear eigenvalue problem. Informed by these observations, we introduce the following scalings
\begin{equation}
H = \dfrac{h\sqrt{\tilde{f}_a}\mathcal{F}_a(U)}{U}, \quad Z=\frac{z}{\ell}, \quad U = \frac{Ca_\kappa u}{\sqrt{ \tilde{f}_a}}, \quad \mathcal{F}_a(U)=\frac{f_a\left(u(U)\right)}{\tilde{f}_a},\label{eq:scaling}
\end{equation}%
where $\tilde{f}_a=\lim\limits_{\ell\rightarrow \infty}f_a\left(u(\ell)\right)$, which is strictly positive by assumption. We note that our following analysis holds more generally than just for the specific function $f_a(u)$ defined in~\eqref{eq:active forcing chemotaxis}. Specifically, it holds for any $f_a$ such that
\begin{equation}
f_a(u)>0, \quad f'_{a}(u)<0 \text{ for all } u\geq0.\label{eq:condition forcing}
\end{equation}
Biologically, this general form~\eqref{eq:condition forcing} corresponds to a faster-moving aggregate producing shallower gradients, thereby reducing the overall driving force for motion. %This captures the natural trade-off that limits the self-generated chemotaxis of cell collectives.

These scalings result in the width of the rescaled outer region being of $\mathcal{O}(1)$, whereas the widths of the rescaled rear and front boundary layers now shrink to zero as $\ell\rightarrow\infty$. 
Using~\eqref{eq:scaling}, the problem \eqref{eq:general_TW_problem} converts into the following nonlinear eigenvalue problem
\begin{subequations}
\begin{align}
\varepsilon^3w(U)^3 H''' =\frac{1}{H}-1, \quad Z\in\left(-\frac{1}{2},\frac{1}{2}\right)\label{eq:outer}
\end{align}
with boundary conditions
\begin{align}
H\left(\pm\frac{1}{2}\right) = 0, \quad
\varepsilon \frac{U}{\mathcal{F}_a(U)} H'\left(\pm\frac{1}{2}\right) = \mp\sqrt{1\pm\mathcal{N}_\theta U}, 
\end{align}\label{problem_constant:elongated}%
\end{subequations}
where 
\begin{equation}
\mathcal{N}_\theta:=\sqrt{\tilde{f}_a}\frac{\eta_\theta}{Ca_\kappa}=\mathcal{O}(1),\quad w(U):=\left(\frac{U}{\mathcal{F}_a(U)^{2}}\right)^{1/3},\label{eq:def_ntheta}
\end{equation}
and 
\begin{equation}
\varepsilon := \frac{1}{\ell \sqrt{\tilde{f}_a}}\ll1. \label{def:varepsilon}
\end{equation}
The large droplet limit $\ell \to \infty$ therefore corresponds to the singular limit $\varepsilon\rightarrow0$, provided that $\varepsilon^3U/\mathcal{F}_a(U)^{2}\ll1$, and $\varepsilon U/\mathcal{F}_a(U)\ll1$. In the remainder of this section, we present the full details of the large-$\ell$ analysis. Readers primarily interested in the physical interpretation of our theory in the context of self-generated chemotaxis can skip ahead to~\Cref{app:self-generated chemotaxis_conditions}. 

\subsection{Leading-order problem}
\label{sec:leading-order problem}
We start by deriving the leading-order solution to the eigenvalue problem~\eqref{problem_constant:elongated}. We find that matching the behaviours between the outer region and the rear boundary layer is sufficient to determine the leading-order travelling speed $U_0$ (see workflow~\Cref{fig:asymptotics work flow}). This is because the leading contribution to the rescaled shape profile $H$ in the rear is set independently of the travelling speed. By contrast, the height profile in the front boundary layer adapts to match the rear behaviour. We will see that the information of the leading-order solution in the front boundary layer is key to determining the exponentially small correction to the travelling speed $U$, which is required to understand the overall morphological transitions of the droplet.

\subsubsection{Solution in the outer region: $Z\pm1/2=\mathcal{O}(1)$}
As outlined in the flow of information in~\Cref{fig:asymptotics work flow}, we start by considering the solution to the leading-order outer problem~\eqref{problem_constant:elongated}. In the outer region, the boundary conditions can be neglected and 
\begin{equation}
H_0\equiv 1
\end{equation}
is an exact solution to \eqref{eq:outer}.
\subsubsection{Solution in the rear boundary layer: $Z+1/2=\mathcal{O}(\varepsilon)$}
\label{sec:rear leading problem}
Next, we consider the leading-order problem in the boundary layer near the rear contact line.  We scale into the rear boundary layer by rescaling the spatial dimension $Z$
\begin{equation}
Z =-\frac{1}{2}+\varepsilon w(U)\sigma, \quad \sigma\in(0,\infty).\label{eq:spatial scaling rear}
\end{equation}
Substituting~\eqref{eq:spatial scaling rear} into~\eqref{problem_constant:elongated}, we find that the height profile $\BLI{H}(\sigma)=H(Z(\sigma))$ satisfies the following leading-order problem 
\begin{subequations}
    \begin{align}
        \BLI{H}_0'''&=\frac{1}{\BLI{H_0}}-1,\quad \sigma\in(0,\infty)\label{eq:Y0 rear}\\
          \lim_{\sigma\rightarrow\infty}\BLI{H}_0(\sigma)&=1,\quad \BLI{H}_0(0)=0, \quad \BLI{H}'_0(0)=\frac{\sqrt{1-\mathcal{N}_\theta U_0}}{U^{2/3}_0}.\label{BCI:leading}
    \end{align}\label{problem constant inner rear leading-order}%
\end{subequations}
The governing equation~\eqref{eq:Y0 rear} is a third-order ODE, and hence it has three degrees of freedom. The conditions at $\sigma=0$ remove two degrees of freedom. The far-field condition in~\eqref{BCI:leading} is equivalent to an additional two independent constraints. This can be seen by a far-field WKBJ analysis (\emph{e.g.},~\cite{bender_wkb_1999,chapman_role_2023}) of~\eqref{eq:Y0 rear} in the limit $\sigma\rightarrow \infty$, which gives 
\begin{equation}
\BLI{H}_0\sim 1+\BLI{A}_0e^{-\sigma}+\BLI{B}_0e^{\sigma/2}\sin\left(\frac{\sqrt{3}\sigma}{2}+\BLI{\psi}_0\right), \quad \sigma\gg 1,\label{eq:farfieldI}
\end{equation}
where $\BLI{A}_0$, $\BLI{B}_0$ and $\BLI{\psi}_0$ are constants of integration.
By imposing that $\BLI{H}_0\to1$ in the far field (\emph{i.e.}, by matching to the outer solution), we must have $\BLI{B}_0=0$, which also removes the $\BLI{\psi}_0$ degree of freedom. Hence, the single far-field condition removes two degrees of freedom. Since~\eqref{BCI:leading} removes four degrees of freedom from the third-order ODE~\eqref{eq:Y0 rear}, we may conclude that the problem~\eqref{problem constant inner rear leading-order} is locally closed and sets the value of the as-of-yet unknown leading-order travelling wave speed $U_0$.

We solve~\eqref{problem constant inner rear leading-order} numerically using a shooting method that exploits the translational invariance of~\eqref{problem constant inner rear leading-order}. We integrate~\eqref{eq:Y0 rear} backwards in $\sigma$ from an arbitrary and large value of $\sigma=\sigma^*\gg1$ (where we set the far-field condition $\BLI{H}_0(\sigma)\sim 1-e^{-\sigma}$) until $\BLI{H}_0=0$. Once we determine where $\BLI{H}_0=0$ from our shooting solution, we translate it so that $\BLI{H}_0(0)=0$ to obtain the solution to~\eqref{problem constant inner rear leading-order}. More details are provided in~\Cref{app: Leading-order solution in the rear boundary layer}. This procedure allows us to estimate the actual value of the constant $\BLI{A}_0\approx -0.84$ and $\BLI{H}_0'(0)\approx1.345$ independently of the model parameters. Using the latter with Eq.~(\ref{BCI:leading}) allows us to deduce that $U_0$ is defined implicitly by the nonlinear equation
\begin{equation}
\left(\frac{U_0}{\mu_0}\right)^{4/3}+\mathcal{N}_\theta U_0-1=0,\label{eq constant: U_0}%
\end{equation}
where the constant $\mu_0=\BLI{H}_0'(0)^{-3/2}\approx 0.641$. Eq. \eqref{eq constant: U_0} uniquely defines a single positive real $U_0$. We also deduce that $U_0\leq\mu_0$ (for physical values of $\mathcal{N}_\theta\geq0$). Although the roots of~\eqref{eq constant: U_0} can be computed explicitly, the complexity of the expression does not provide insight; a plot of the relationship between $U_0$ and $\mathcal{N}_\theta$ is shown in~\Cref{fig:regionIA}. We note instead that for parameter choices such that $\mu_0\mathcal{N}_\theta>1$, the dissipation of energy at the contact lines becomes the dominant limiting factor for the droplet speed, \emph{i.e.}, $U_0<\min\left\{\mathcal{N}_{\theta}^{-1},\mu_0\right\}$. Energy dissipation at the contact line and the resulting adjustment of the contact angle reduce the speed at which the droplet migrates as it generates an effective force that opposes the constant active forcing driving migration.
\subsubsection{Solution in the front boundary layer: $Z-1/2=\mathcal{O}(\varepsilon)$}
Given the behaviour at the rear of the droplet, we can now compute the droplet height profile in the front boundary layer. We resolve the solution near the boundary layer at the front contact line by rescaling the spatial dimension $Z$
\begin{equation}
Z =\frac{1}{2}+\varepsilon\sigma w(U) , \quad \sigma\in(-\infty,0).\label{eq:spatial scale front}
\end{equation}
Substituting~\eqref{eq:spatial scale front} into~\eqref{problem_constant:elongated}, we find that the leading-order behaviour of the height profile $\BLII{H}(\sigma)=H(Z(\sigma))$ satisfies the problem
\begin{subequations}
    \begin{align}
        \BLII{H}_0'''&=\frac{1}{\BLII{H}_0}-1,\quad \sigma\in(-\infty,0)\label{eq:Y0 leading front}\\
       % \BLII{H}_0&\sim 1+\BLII{A}_0e^{-\sigma}+\BLII{B}_0e^{\sigma/2}\sin\left(\frac{\sqrt{3}\sigma}{2}+\BLII{\psi}_0\right), \quad \sigma\rightarrow-\infty\label{eq:farfield front}\\
          \lim_{\sigma\rightarrow-\infty}\BLII{H}_0(\sigma)&=1,\quad\BLII{H}_0(0)=0, \quad \BLII{H}_0'(0)=-\frac{1}{\mu_0^{2/3}}\sqrt{\frac{2\mu^{4/3}_0}{U^{4/3}_0}- 1},\label{front leading-order BC}
    \end{align}\label{problem constant inner LHS leading-order}%
\end{subequations}
where $\mu_0$ and $U_0$ are both known positive constants. The far-field behaviour of~\eqref{eq:Y0 leading front} around $H\equiv 1$ has the same functional form as~\eqref{eq:farfieldI}
\begin{equation}
     \BLII{H}_0\sim 1+\BLII{A}_0e^{-\sigma}+e^{\frac{\sigma+\BLII{\sigma}_0}{2}}\sin\left(\frac{\sqrt{3}(\sigma+\BLII{\sigma}_0)}{2}+\BLII{\psi}_0\right), \quad \sigma\rightarrow-\infty,\label{eq:farfield front}
\end{equation}
using $\BLII{B}_0=\exp(\BLII{\sigma}_0/2)$ for later convenience, and with the difference that now $\sigma\to -\infty$. Therefore, imposing that $\BLII{H}_0$ is bounded in the far field as $\sigma\rightarrow -\infty$ now only sets one degree of freedom ($\BLII{A}_0=0$). Hence, given that $U_0$ is now known, the far field condition with the two additional conditions at $\sigma=0$ close~\eqref{problem constant inner LHS leading-order}. The key quantities we will require from this inner region are the phase shift $\BLII{\psi_0}$ and the scaling factor $\BLII{\sigma}_0$. To obtain these, we again use the shooting method to obtain the solution $\BLII{H}_0$ as a function of $\mathcal{N}_\theta$ (see~\Cref{fig:regionID}) and its far-field behaviour (see Figures~\ref{fig:regionIB}-\ref{fig:regionIC}). 
We present the details of the numerical procedure used to approximate $\BLII{H}_0$, and the far-field constant $\BLII{\sigma}_0$ and $\BLII{\psi}_0$ (\Cref{fig:regionIC}) in~\Cref{app: Leading-order solution in the front boundary layer}. We note that, in the limit of $U_0\rightarrow 0$, corresponding to $\mathcal{N}_\theta\gg1$ (\Cref{fig:regionIA}), the problem becomes singular since $\BLII{H}_0'(0)\rightarrow \infty$. This corresponds to the situation in which the contact angle at the rear of the droplet approaches zero. By contrast, the front contact angle approaches $\pi/2$, and the net force generated by the difference in contact angles counteracts the active forcing experienced by the droplet. We find numerically that this occurs for a far field phase shift $\BLII{\psi}_0^{cr} \approx -3\pi/4$.

\subsection{First-order correction}
The leading-order approximation as $\ell\to\infty$ is not sufficient to explain proliferation-driven morphological transitions in migrating droplets. As we will see, this is because the transition in TW morphologies and the impact of droplet size on the migration speed depend on finite-size effects. To properly account for such finite-size effects, we must resolve how the exponentially small terms originating at the front boundary layer propagate into the rear boundary layer. 
\subsubsection{Solution in the outer region: $Z\pm1/2=\mathcal{O}(1)$}
As outlined in the flow of information in~\Cref{fig:asymptotics work flow}, we return to the outer problem~\eqref{problem_constant:elongated}, but now considering higher-order terms.
In the outer region, the leading-order solution $H_0\equiv 1$ is an exact solution to the problem~\eqref{eq:outer}. Hence, any perturbation to the solution generated by the boundary conditions must be exponentially small. Importantly, the mechanisms that determine the admissible travelling wave solutions depend on these exponentially small effects that originate near boundaries. Hence, classic Poincaré asymptotic series in powers of $\varepsilon$ do not suffice to approximate the solution, and the exponentially small terms must be included. Again, we find the correction by linearising Eq.~\eqref{eq:outer} around the leading-order solution, posing $H\sim 1+H_1$. We obtain:
\begin{equation}
    H_1 =A^{\dagger}\exp\left[-\frac{Z+1/2}{w(U)\varepsilon}\right]+B^{\dagger}\exp\left[\frac{Z-1/2}{2w(U)\varepsilon}\right]\sin\left(\frac{\sqrt{3}(Z-1/2)}{2w(U)\varepsilon}+\psi^{\dagger}\right)\label{eq:outer_H}
\end{equation}
where $A^{\dagger}$, $B^{\dagger}$ and $\psi^\dagger$ are unknown constants. Although both exponential terms in~\eqref{eq:outer_H} are small in the outer region, each matches with leading-order terms in the boundary layers. This is apparent when expressing $H$ in terms of the inner variable for the rear
\begin{equation}
\begin{aligned}
    H_1(\sigma)\sim A^\dagger e^{-\sigma} + \exp\left(-\frac{1}{2\varepsilon w(U)}\right)B^\dagger e^{\sigma/2} &\sin\left(\frac{\sqrt{3}}{2}\left(\sigma-\frac{1}{\varepsilon U_0^{1/3}}\right)+\psi^{\dagger}\right)
    \end{aligned}, \quad \sigma \rightarrow \infty\label{far field outer rear}
\end{equation}
and front boundary layers
\begin{equation}
    H_1(\sigma)\sim B^\dagger e^{-\sigma/2}\sin\left(\frac{\sqrt{3}\sigma}{2}+\psi^\dagger\right) +\mathcal{O}\left(\exp\left(-\frac{1}{\varepsilon w(U)}\right)\right), \quad \sigma \rightarrow -\infty.\label{far field outer front}
\end{equation}
By matching~\eqref{far field outer rear}-\eqref{far field outer front} with the far-field behaviour of the height profile respectively in the rear~\eqref{eq:farfieldI} and front~\eqref{eq:farfield front} boundary layers, we can determine the unknown constants $A^\dagger$, $B^\dagger$, and $\psi^\dagger$:
\begin{equation}
    A^\dagger=\BLI{A}_0, \quad B^{\dagger}=e^{\BLII{\sigma}_0/2},\quad \psi^{\dagger}=\BLII{\psi}_0+\frac{\sqrt{3}\BLII{\sigma}_0}{2},\label{matching outer front BL}
\end{equation}
where we recall $\BLI{A}_0$ is a known constant independent of model parameters, while $\BLII{\sigma}_0$ and $\BLII{\psi}_0$, and, therefore, $B_0^{\dagger}$ and $ \psi_0^{\dagger}$, depend on the parameter $\mathcal{N}_\theta$ as shown in Figures~\ref{fig:regionIB}-\ref{fig:regionIC}.

Comparing Eqs.~(\ref{far field outer rear})-(\ref{far field outer front}), we find that the leading exponential correction is of order $\mathcal{O}\left(\exp[-1/2\varepsilon w(U)]\right)$. Using this insight, we proceed by expanding all remaining variables, namely the travelling speed $U$ and the height profile in the rear and front boundary layers, as an exponential series expansion
\begin{equation}\label{eq:general exponential expansion}
\varphi\sim \varphi_0+\varphi_1\exp\left[-\frac{1}{2U_0^{1/3}\varepsilon}\right], \quad \varphi\in\left\{\BLI{H},\BLII{H},U\right\}
\end{equation}
where the leading-order contributions are as derived in~\Cref{sec:leading-order problem} and the choice of exponential decays corresponds to the slower decay mode obtained via the WKBJ approximation~\eqref{eq:outer_H}. The $U_0^{1/3}$ term in the denominator comes from the leading-order contribution of $w(U)$~\eqref{eq:def_ntheta}, which is obtained by evaluating $w(U_0)=U_0^{1/3}$ at the leading-order speed $U_0$, which is now known.
\subsubsection{Solution in the rear boundary layer: $Z+1/2=\mathcal{O}(\varepsilon)$}
To determine $U_1$, we must first resolve the height profile $\BLI{H}_1$ in the boundary layer near the rear contact line, which satisfies the following non-autonomous boundary value problem
\begin{subequations}\label{eq:Y1 rear}
    \begin{align}
       \BLI{H}'''_1+\frac{\BLI{H}_1}{\BLI{H}_0^2(\sigma)}&=0,   \label{eq first correction rear}  \\
        \BLI{H}_1(0)&=0, \quad \BLI{H}_1'(0)=-\left(\frac{4-\mathcal{N}_\theta U_0}{1-\mathcal{N}_\theta U_0}-2U_0\mathcal{F}'_{\mathcal{A}}(U_0)\right)\frac{U_1}{6 U_0\mu^{2/3}_0},\label{BC first correction rear}\\
        \BLI{H}_1&\sim B^\dagger e^{\sigma/2}\sin\left(\frac{\sqrt{3}}{2}\left(\sigma-\frac{1}{\varepsilon U_0^{1/3}}\right)+\psi^{\dagger}\right), \quad \sigma\rightarrow\infty\label{far field first correction rear}
    \end{align}
\end{subequations}
where the far-field condition is obtained by matching the exponentially small correction to the outer solution~\eqref{far field outer rear}. Similar to the leading-order problem in this boundary layer (see~\Cref{sec:rear leading problem}), the single far-field condition~\eqref{far field first correction rear} imposes two, rather than one, independent constraints, which can be shown via a WKBJ analysis of the far-field behaviour of~\eqref{eq first correction rear}. This, together with the two constraints at $\sigma=0$~\eqref{BC first correction rear}, overdetermines the third-order ODE~\eqref{eq first correction rear}, implying that a solution exists only for a specific value of the unknown speed correction $U_1$. Exploiting the linearity of the system \eqref{eq:Y1 rear}, we can solve~\eqref{far field first correction rear} via the shooting method (see~\Cref{app:computation BL}) and obtain an explicit expression for $U_1$ uniquely in terms of the leading-order solutions 
\begin{equation}
U_1(\varepsilon)= \frac{6 U^{7/3}_0 \mu_0^{2/3}B^\dagger_0|\BLI{ s}|}{U_0^{4/3}+3\mu_0^{4/3}-2U_0^{7/3}\mathcal{F}'_{\mathcal{A}}(U_0)}\sin\left(\frac{\sqrt{3}}{2 U_0^{1/3}\varepsilon}-\psi^\dagger-\arg \BLI{s}\right).\label{eq:U1}
\end{equation}
Here, the constant $\BLI{s}\in\mathbb{C}$ is uniquely defined by the problem~\eqref{eq:Y1 rear} independently of the model parameters; its numerically approximated value is $\BLI{s}\approx 0.824 + 0.106 i$.
In~\eqref{eq:U1}, $|\cdot|$ and $\arg\cdot$ indicate the modulus and argument of a complex number. Details on how~\eqref{eq:U1} is determined from~\eqref{eq:Y1 rear} can be found in~\Cref{app:computation BL}. We note that, as long as $\mathcal{F}_a'(U_0)<0$ for all $U_0>0$ (such as for the biologically relevant scenario considered here), the denominator of Eq.~(\ref{eq:U1}) is always positive and, therefore, $U_1$ remains finite for any finite value of $U_0$. 

To fully determine the height profile at next order, we can construct a solution for the first correction to the exponential expansion in the boundary layer at the front contact line $\BLII{H_1}$ (see~\Cref{app:computation BL}); while the solution allows us to gather insights on the behaviour of the droplet near the front, resolving it is not necessary in order to characterise the emergent migration dynamics of the droplet. 

\section{An asymptotic theory of proliferation-driven morphological transitions in chemotactic active droplets}
\label{app:self-generated chemotaxis_conditions}

To briefly summarise, in~\Cref{sec logathmic asymptotics}, we considered asymptotic solutions up to a first-order correction to the non-linear travelling wave problem~\eqref{eq:general_TW_problem} in the elongated-droplet limit; namely, the singular limit $\varepsilon\to0$ in the rescaled problem~\eqref{problem_constant:elongated}. Our analysis yielded asymptotic approximations for the morphology and speed of droplets, including the expressions~\eqref{eq constant: U_0} and~\eqref{eq:U1} for the leading- and first-order terms in the expansion of the rescaled migration speed $U$~\eqref{eq:general exponential expansion}. The obtained expression captures the non-monotonic relationship between the droplet volume and speed in Figure~\ref{fig bifurcation constant forcing}, and its derivation highlights how the maximal migration speed of droplets is set by the information propagation (i.e. asymptotic matching) of exponentially small corrections in the droplet height originating near the contact lines.

In this section, our goal is to use the results from~\Cref{sec logathmic asymptotics} to gain insights into the physics that drives the proliferation-driven morphological transitions seen in Figures
\ref{fig:SG_travelling_homogeneous_example}-\ref{fig bifurcation constant forcing} and the mechanisms that determine their nature (\emph{i.e.}, smooth vs sharp transitions).

\subsection{Physical characterisation of elongated droplets' migration}
By definition of $\varepsilon$~\eqref{def:varepsilon}, the limit $\varepsilon\to0$ corresponds to a regime in which the droplet length $\ell$ is much larger than the asymptotic capillary length $\tilde{\ell}_\kappa=1/\sqrt{\tilde{f}_a}$, where we recall that $\tilde{f}_a$ is the asymptotic value of the active forcing~\eqref{eq:active forcing chemotaxis} as $\ell \to\infty$. Physically, this is a regime in which there can be no global balance between activity and surface tension across the droplet. We parameterise the elongated travelling wave solutions to~\eqref{problem_constant:elongated} in terms of the non-dimensional variable
\begin{equation}
    L:=\frac{1}{\varepsilon}=\frac{\ell}{\tilde{\ell}_\kappa}\gg 1,\label{def L}
\end{equation}
which measures the length of the droplet relative to the asymptotic capillary length $\tilde{\ell}_\kappa$. We find that the leading-order contribution to the non-dimensional speed $U_0$ is independent of $L$ and set by the local behaviour of the droplet profile near the rear contact line. The constant $U_0$ instead depends uniquely on the non-dimensional parameter $\mathcal{N}_\theta\geq0$ via~(\ref{eq constant: U_0}) and spans values in the range $(0,\mu_0]$ (see~\Cref{fig:regionIA}). To gain physical insight into~(\ref{eq constant: U_0}), we write it in terms of the non-dimensional parameter
\begin{equation}
\gamma_\theta:=(1-\eta_\theta u_0)^{3/4}\in(0,1],\label{eq:gammatheta}
\end{equation}
which quantifies the extent to which the asymptotic droplet migration is limited by contact line dissipation. Then, $\mathcal{N}_\theta$ can be rewritten as
\begin{equation}
\mathcal{N}_\theta=\frac{1-\gamma_\theta^{4/3}}{\mu_0\gamma_\theta},\label{eq:Ntheta_gamma_theta}
\end{equation}
highlighting how $\gamma_\theta\to1$ corresponds to negligible contact line dissipation ($\mathcal{N_\theta}\rightarrow0$), while $\gamma_\theta\to0$ indicates a regime in which contact line dissipation dominates ($\mathcal{N_\theta}\gg1$).
Using~\eqref{eq:scaling},~\eqref{eq:def_ntheta} and ~\eqref{eq:gammatheta}, we can rewrite~\eqref{eq constant: U_0} as 
\begin{equation}
    \gamma_\theta=\frac{Ca^0_\kappa \tilde{\ell}_\kappa}{\mu_0},\label{eq:asymptotic u_0}
\end{equation}%
where $\mu_0\approx0.641$ is a known constant obtained from the leading-order solution (see~\Cref{sec:rear leading problem}), $Ca_\kappa^0=Ca_\kappa u_0$ is the asymptotic capillary number, and $\tilde{\ell}_\kappa$ is the asymptotic capillary length noted above. Combining~\eqref{eq:gammatheta}-\eqref{eq:asymptotic u_0}, we can interpret $\mu_0$ as the maximum value that the product $Ca^0_\kappa\tilde{\ell}_\kappa$ can attain (occurring when contact line dissipation is neglected). This product represents the characteristic length scale over which viscous stresses deform the interface before being counteracted by active forcing and surface tension. Accordingly, Eq.~\eqref{eq:asymptotic u_0} relates contact line dissipation to the extent to which viscous stresses shape the morphology of elongated droplets. Small values of $\gamma_\theta$ (namely, large contact line dissipation) are associated with droplets in which the viscous stresses have a minimal role in shaping the morphology of the droplet. By contrast, values of $\gamma_\theta\sim 1$ result in elongated droplets where the length scale over which viscous-capillary-active flows balance is $\mathcal{O}(1)$, and, therefore, viscous stresses play a significant role in shaping the droplet morphology.

\begin{figure}
    \centering
           \begin{subfigure}{0.00\textwidth}
    \captionlistentry{}
    \label{fig: approximate theory speed A}
    \end{subfigure}
               \begin{subfigure}{0.0\textwidth}
    \captionlistentry{}
    \label{fig: approximate theory speed B}
    \end{subfigure}
    \includegraphics[width=0.85\linewidth]{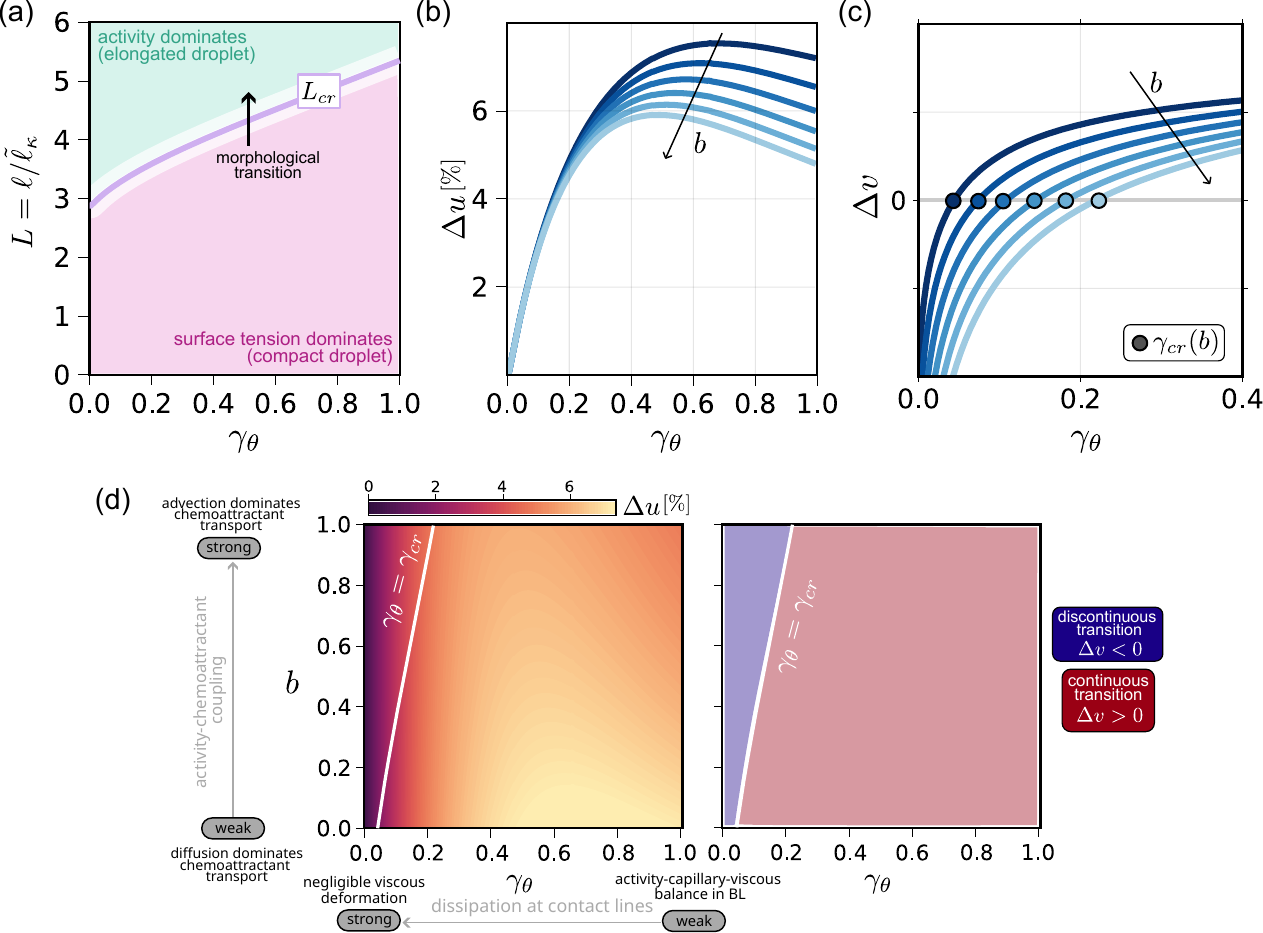}
               \begin{subfigure}{0.00\textwidth}
    \captionlistentry{}
    \label{fig: approximate theory mass A}
    \end{subfigure}
               \begin{subfigure}{0.00\textwidth}
    \captionlistentry{}
        \label{fig: approximate theory mass B}
    \end{subfigure}
    \caption{Elongated-droplet theory of morphological transitions. Theoretical predictions for how (a) the critical length $L_{cr}(\gamma_\theta)$ at which $u$~\eqref{eq leading-order function of L} attains its maximal value, (b) the drop in migration speed $\Delta u(\gamma_\theta,b)$~\eqref{eq:Delta u asymptotic}, and (c) the volume variation $\Delta v(\gamma_\theta,b)$~\eqref{eq:Delta v} depend on the non-dimensional parameter $\gamma_\theta$~\eqref{eq:gammatheta}. In (b)-(c), colors indicate different values of the non-dimensional parameter $b$~\eqref{eq:definition b}, which ranges from $b=0$ (dark blue) to $b=1$ (light blue).  (d) Phase diagram of elongated droplets undergoing self-generated chemotaxis predicted by the asymptotic theory. The two parameters $\gamma_\theta$ and $b$ characterise the balance between the different physical mechanisms governing the migration of elongated droplets (more details in the main text). In the left panel, the colormap indicates the relative reduction in the migration speed $\Delta u$ (defined by Eq.~\eqref{eq:Delta u asymptotic}) following morphological transitions. The white curve indicates the value of the parameter for which $\Delta v=0$, partitioning the region of parameter space where the theory predicts proliferation to drive a continuous (blue shaded area) or a discontinuous (red shaded area) morphological transition.}
    \label{fig: approximate theory speed}
\end{figure}
Although the leading-order speed $u_0$ is constant in $L$ in the large-droplet limit,~\eqref{def volume dimensional} reveals that the droplet volume increases linearly with $L$
\begin{equation}
    \frac{v_0}{(\tilde{\ell}_\kappa)^2}= \mu_0 \gamma_\theta L+ 1-\gamma_\theta^{4/3}.\label{eq:volume leading-order}
\end{equation}
Eq.~(\ref{eq:volume leading-order}) highlights how $\gamma_\theta$ mediates the sensitivity of the droplet volume to changes in its length. In particular, small values of $\gamma_\theta$, corresponding to large dissipation at the contact line, imply that even substantial increases in droplet length lead to only small variations in droplet volume. This behaviour is consistent with the results shown in~\Cref{fig:SG_travelling_homogeneous_example,fig bifurcation constant forcing}, where a stronger contact-line dissipation ($\gamma_\theta\approx0.03$) gives rise to the formation of a thin trailing layer whose volume is negligible compared to the mass accumulating at the droplet front.
\subsection{Physical characterisation of proliferation-driven morphological transitions}
The analysis in~\Cref{sec logathmic asymptotics} reveals that the structure of the leading-order solution to~\eqref{problem_constant:elongated} is locally selected by the behaviour near the rear of the droplet. By contrast, we find that the corresponding corrections capture the global shape of the droplet height and account for finite-size effects. In particular, this depends on the matching of the exponentially small corrections arising within the front boundary layer with the solution in the middle $\mathcal{O}(1)$ region (\Cref{fig:schematic asympotics}). 
Substituting~\eqref{eq:U1} into~\eqref{eq:general exponential expansion}, we obtain the following scaling law for the droplet speed
\begin{equation}
\frac{u(L;\gamma_\theta,b)}{u_0}\sim 1+\frac{g(\gamma_\theta)\exp\left[-\frac{L-L_{cr}(\gamma_\theta)}{2(\mu_0\gamma_\theta)^{1/3}}\right]}{1+\frac{2b\gamma_\theta^{4/3}}{\gamma_\theta^{4/3}+3}}\sin\left(\frac{\sqrt{3}\left(L-L_{cr}(\gamma_\theta)\right)}{2(\mu_0\gamma_\theta)^{1/3}}+\frac{\pi}{3}\right),\label{eq leading-order function of L}\end{equation}
which depends on $\gamma_\theta$ and the non-dimensional parameter
\begin{equation}
b=-\mathcal{F}'_{\mathcal{A}}(U_0)U_0=\frac{-f_a'(u_0)u_0}{f_a(u_0)}=\frac{\eta_c u_0}{\sqrt{\eta_c^2u^2_0+4}}\in[0,1),\label{eq:definition b}
\end{equation} 
which characterises the strength of the coupling between the migration speed and the active forcing. In Eq.~\eqref{eq:definition b}, $\Pe^0_c:=\eta_cu_0>0$ is the asymptotic chemoattractant P\'{e}clet number. We recall that $\Pe^0_c\gg1$ (equivalently $b\to1$) corresponds to the regime in which the transport of the chemoattractant is limited by the migration speed. This, therefore, corresponds to a regime of strong coupling between the migration dynamics (\emph{i.e.}, speed) and the active forcing (\emph{i.e.}, chemoattractant gradients). Conversely, $\Pe^0_c\ll1$ (equivalently $b\to0$) indicates the regime in which the primary mechanism of chemoattractant transport (in the travelling wave reference frame) is diffusion, and, therefore, there is a weak coupling between the migration dynamics and the active forcing strength. 

Using Eq.~\eqref{eq leading-order function of L}, we are now in a position to use the asymptotic results to capture the morphological transitions observed in~\Cref{fig bifurcation constant forcing}.  Although Eq.~\eqref{eq leading-order function of L} is formally only valid for large $L$ (\emph{i.e.}, exponentially small first-correction), we find excellent agreement with the results of the numerical bifurcation analysis of~\eqref{eq:general_TW_problem} for even moderate values of $L$. In particular, the asymptotic result accurately captures the regime $L\approx L_{cr}$ where the droplet transitions from a compact to an elongated morphology (Figure~\ref{fig bifurcation constant forcing}). Eq.~\eqref{eq leading-order function of L} reveals the oscillatory dependence of the speed $u$ on $L$, originating from the first-order correction. This oscillatory behaviour is key to capturing the maximal migration speed and, more generally, how droplet size limits its migration. 
The function $L_{cr}(\gamma_\theta)$ defines the critical value of the ratio $\ell/\tilde{\ell}_\kappa$ at which the droplet attains its maximal speed and the droplet starts its transition from a compact to an elongated morphology due to the change in the balance between surface tension and activity. As shown in~\Cref{fig: approximate theory speed A}, $L_{cr}$ monotonically increases with $\gamma_\theta$. This can be rationalised by examining the behaviour of the solution near the front contact line as the contribution of contact line dissipation increases (\Cref{fig:regionID}). A smaller value of $\gamma_\theta$ (\emph{i.e.}, larger value of $\mathcal{N}_\theta$) results in a more asymmetric distribution of the mass, with a larger fraction accumulating at the front rather than the rear. This implies a more significant difference in the frictional forces at the front and rear of the droplet that eventually facilitates the transition to an elongated morphology.

We now use the asymptotic theory to investigate the impact of morphological transitions on the droplet speed. Although $L_{cr}$ is modulated by the strength of the activity-capillary-viscous balance (via $\gamma_\theta$), it is independent of the coupling between migration and the chemoattractant gradients (namely $b$). By contrast, the amplitude of the speed fluctuations depends on both $b$ and $\gamma_\theta$. We quantify this by looking at the predicted reduction in the droplet speed associated with the transition to an elongated morphology
\begin{equation}
\Delta u(\gamma_\theta,b)=\frac{u(L_{cr}(\gamma_\theta))}{u_0}-1\sim\frac{\sqrt{3}}{2}\dfrac{g(\gamma_\theta)}{1+\dfrac{2b\gamma_\theta^{4/3}}{\gamma_\theta^{4/3}+3}},\label{eq:Delta u asymptotic}
\end{equation}
where $g$ is a known positive function~\eqref{app:eq g} describing the amplitude of the fluctuations in the droplet speed in the absence of any coupling to the chemoattractant gradients (\emph{i.e.}, $b=0$). Since $g$ depends on the behaviour of the first-order correction to the droplet profile in the rear droplet boundary layer, which itself is a function of $\gamma_\theta$, its functional form cannot be obtained explicitly. Yet, we can estimate it numerically using the approximate shooting solution to~\eqref{eq:H_first_order_correction} (see \Cref{fig: app g} in~\Cref{computation g}). We find that $g$ has a non-monotonic dependency on $\gamma_\theta$. From~\eqref{eq:Delta u asymptotic}, it is apparent that increasing $b$ reduces the drop in the migration speed associated with proliferation-driven morphological transitions. The dependence of $\Delta u$ on $\gamma_\theta$ is less straightforward. Estimating $\Delta u$ numerically, we find that $\Delta u$ increases from zero at small $\gamma_\theta$ up to a maximal value, beyond which any increase in $\gamma_\theta$ results in a reduction of $\Delta u$ (\Cref{fig: approximate theory speed B}). We conclude that the coupling between migration and chemoattractant gradients, quantified by $b$, can have a significant impact on the changes in migration dynamics induced by the transition to an elongated droplet morphology. In contrast to variations in $\gamma_\theta$, changing $b$ regulates the sensitivity of the collective migration dynamics to droplet size without altering the internal stress balance within the droplet.

We conclude by employing our asymptotic theory to characterise the nature of proliferation-driven morphological transitions during self-generated chemotaxis across the whole of parameter space. As mentioned at the beginning of~\Cref{sec logathmic asymptotics}, mathematically, the length of the droplet is the most natural choice of the control parameter to study the behaviour of TW solutions of our model. However, in the biological context, the volume, rather than the length of the droplet, is the quantity that is regulated via proliferation. We can use Eq.~\eqref{eq leading-order function of L} and~\eqref{def volume dimensional} to obtain an approximate relation between the droplet volume $v$ and its length $L$ up to a first-order correction
\begin{equation}
\begin{aligned}
    \frac{v}{\mu_0(\tilde{\ell}_\kappa)^2}\sim &\left(\gamma_\theta L+\frac{1-\gamma_\theta^{4/3}}{\mu_0}\right)\left[1
    +\frac{(1+b)g(\gamma_\theta)e^{-\frac{L-L_{cr}(\gamma_\theta)}{2(\mu_0\gamma_\theta)^{1/3}}}}{1+\frac{2b\gamma_\theta^{4/3}}{\gamma_\theta^{4/3}+3}}\sin\left(\frac{\sqrt{3}\left(L-L_{cr}(\gamma_\theta)\right)}{2(\mu_0\gamma_\theta)^{1/3}}+\frac{\pi}{3}\right)\right].\label{eq:volume up to first-order correction}
    \end{aligned}
\end{equation}
When $v(L)$ is monotonic, Eq.~\eqref{eq:volume up to first-order correction} can be inverted, implying that travelling wave solutions may be equivalently parametrised by either their length or their volume. By contrast, if $v(L)$ is non-monotonic, travelling wave solutions with different lengths are mapped to the same volume (as in the example illustrated in~\Cref{fig bifurcation constant forcing panel B}). Inspecting Eq.~\eqref{eq:volume up to first-order correction} reveals that while volume and length are positively correlated for $L\gg1$, for finite values of $L$, the exponential correction (second term in the square brackets) could reverse this trend, resulting in a non-monotonic relationship between droplet length and volume. This correction is proportional to the correction in the speed $u$ (see~\eqref{eq leading-order function of L}), with an additional prefactor $(1+b)$, which controls its overall magnitude. Consequently, while increasing $b$ attenuates oscillations in the droplet speed (Figures~\ref{fig: approximate theory speed B} and~\ref{fig: app velocity} in~\Cref{computation g}), it simultaneously amplifies fluctuations in the volume, favouring non-monotonic $v(L)$ profiles. This trend is verified by plotting the $v(L)$ profile for different values of $b$ (see~\Cref{fig: app volume} in~\Cref{computation g}). Yet, the impact of $b$ on the monotonicity of $v(L)$ is also influenced by the value of $\gamma_\theta$. We further investigate the combined effect of $\gamma_\theta$ and $b$ on the monotonicity of $v(L)$ by evaluating numerically its (normalised) minimal slope
\begin{equation}
    \Delta v(\gamma_\theta,b) = \min\limits_{L\geq L_{cr}(\gamma_\theta)}\frac{v'(L)}{\mu_0\gamma_\theta(\tilde{\ell}_\kappa)^2}.\label{eq:Delta v}
\end{equation}
Since $v'(L)\rightarrow 1$ for $L\to \infty$, $\Delta v<0$ indicates parameter regimes for which $v(L)$ is non-monotonic and, consequently, proliferation-driven transitions are discontinuous (\Cref{fig bifurcation constant forcing panel B}).
When plotting~\eqref{eq:Delta v} as a function of $\gamma_\theta$, we find that $\Delta v$ becomes negative only for sufficiently small $\gamma_\theta<\gamma_{\theta}^{cr}$ (see~\Cref{fig: approximate theory mass A}). Hence, the theory predicts that discontinuous proliferation-driven transitions are only possible when the dissipation at the contact line is sufficiently large to disrupt the local viscous-capillary-active stress balance, so that viscous deformations of the droplet dynamics are negligible. Increasing $b$ enlarges the range of $\gamma_\theta$ values for which the length–volume relation becomes non-monotonic (\Cref{fig: app volume} in~\Cref{computation g}), thereby allowing discontinuous morphological transitions beyond the regime where contact-line dissipation alone dominates the droplet dynamics.

The phase diagram presented in~\Cref{fig: approximate theory mass B} summarises the results of the formal asymptotic analysis, combining the insights obtained from studying the length-velocity~\eqref{eq leading-order function of L} and length-volume~\eqref{eq:volume up to first-order correction} relations. These results are consistent with the numerical bifurcation analysis presented in~\Cref{sec:numerical continuation}, while substantially extending it by enabling a comprehensive characterisation of how model parameters influence the nature of proliferation-driven morphological transitions. The curve $\gamma_{cr}(b)$ partitions the $(\gamma_\theta,b)$-space into two regions (\Cref{fig: approximate theory mass B}) corresponding to continuous ($\Delta v>0$) and discontinuous ($\Delta v<0$) proliferation-driven transitions, revealing how the coupling between chemoattractant and migration (larger $b$) facilitates discontinuous morphological transitions. Looking at the reduction in the droplet speed associated with the transitions, quantified by the metric $\Delta u$~\eqref{eq:Delta u asymptotic}, we find this is reduced in the regime where discontinuous transitions occur, except for very strong migration-chemoattractant coupling. Since $b=0$ corresponds to the gravity-driven passive droplet model proposed in~\citep{peschka_variational_2018}, we find that, in passive droplets, discontinuous transitions manifest only as a sharp jump in droplet length. By contrast, in active droplets guided by self-generated chemotaxis ($b\to1$), such transitions can be significantly different, as they can result in significant changes in both droplet length and migration speed.

\section{Summary and conclusions}
In this work, we study a minimal model of self-generated chemotaxis in growing active droplets, motivated by recent studies of collective chemotaxis in cell groups. The model takes the form of a modified thin-film equation in which the active pressure that drives droplet migration is coupled to the migration speed through an externally produced chemical signal, and where droplet mass increases due to proliferation.

Numerical simulations show that droplet growth, which captures cell proliferation, regulates both droplet morphology and migration speed, driving a transition from a compact and fast-migrating to an elongated and slow-migrating droplet (\Cref{fig:SG_travelling_homogeneous_example}). In~\Cref{sec:main multiple scales}, we derive a reduced model by exploiting the slow timescale of proliferation compared to the timescale of perturbation relaxation in the droplet height profile. This enables a systematic analysis of morphological transitions in terms of changes in the structure of travelling wave solutions as a function of the slowly evolving droplet volume. In this way, proliferation-driven morphological transitions are reduced to understanding how solutions of a nonlinear generalised eigenvalue problem for travelling waves~\eqref{eq:general_TW_problem} depend on the droplet volume.

In~\Cref{sec:numerical continuation}, we first solve the travelling wave problem~\eqref{eq:general_TW_problem} using numerical continuation. The resulting bifurcation diagrams relate the droplet morphology, its speed and length, to its volume $v$, and faithfully reproduce the transition from compact to elongated morphologies observed in direct numerical simulations of the full model (\Cref{fig bifurcation constant forcing}). This analysis shows that two different types of proliferation-driven transitions are possible: (i) continuous transitions, which occur when~\eqref{eq:general_TW_problem} admits a unique solution for each value of $v$; and (ii) discontinuous transitions, which occur when~\eqref{eq:general_TW_problem}  admits multiple solutions for some values of $v$.

In~\Cref{sec logathmic asymptotics}, we use asymptotic techniques to approximate solutions of~\eqref{eq:general_TW_problem}. This analysis reveals that the morphological transitions are a beyond-all-orders phenomenon (\Cref{fig:schematic asympotics}) that requires tracking the evolution of exponentially small perturbations as the droplet length $\ell\to\infty$. Our asymptotic analysis characterises the mathematical structure governing proliferation-driven morphological transitions, and also provides mechanistic insight into the physical processes that regulate their onset and nature. As summarised in~\Cref{fig: approximate theory speed A}, morphological transitions are associated with a change in the balance of capillary and active stresses. Specifically, they are associated with the shift from capillary-dominated stresses in small droplets to activity-dominated stresses in elongated droplets. The disruption of this balance occurs at a critical length, which is controlled by the strength of dissipation at the contact line (measured via the non-dimensional parameter $\gamma_\theta$~\eqref{eq:gammatheta}), which in turn modulates the speed of the droplet and, therefore, the extent to which viscous stresses contribute to shape the droplet morphology. Our formal analysis further allows us to determine the phase diagram of proliferation-driven morphological transitions during self-generated chemotaxis (\Cref{fig: approximate theory mass B}). This reveals how the nature of proliferation-driven morphological transitions is regulated by the interplay of the physical stress balance in the droplet (via $\gamma_\theta$) and the strength of the coupling between the migration speed and chemoattractant gradients (via the non-dimensional parameter $b$).

Overall, we provide a complete characterisation of a minimal active droplet model of collective migration of cell aggregates,  motivated by the experimental system studied in~\cite{ford_pattern_2024}. Our findings reveal how droplet volume limits its migration speed and shapes the morphology of cell aggregates by influencing the balance of passive and active stresses during migration; namely, from capillary-dominated stresses in small fast-moving droplets to activity-dominated stresses in slow-moving elongated droplets. This implies that growth of the aggregate beyond a critical size can be disadvantageous and drive a substantial slowing down of the migration process. Given the generality of our framework, our results have broader implications for the study of the migration of dense cell collectives that display fluid-like properties. Although our minimal model successfully recapitulates some of the key features of how proliferation influences the morphodynamics of active droplets observed in~\citep{ford_pattern_2024}, it necessarily neglects a number of important aspects, such as the possibility of spatial heterogeneity in the active forcing function $f_a$. This could explain the more drastic change in droplet morphology and speed reported in experimental and theoretical investigations of the system, which eventually drives the splitting of the moving aggregate. As such, this study represents a first step towards a theoretical and mathematical understanding of the emergent fluid mechanics of migrating cell aggregates.

\vspace{3mm}
\noindent\textbf{Acknowledgements:} For the purpose of open access, the author has applied a CC-BY public copyright licence to any author-accepted manuscript version arising from this submission. P.P. was supported by a UK Research and Innovation (UKRI) Future Leaders Fellowship (MR/V022385/1).

\noindent\textbf{Data and Software Availability:}
All the data and code necessary to reproduce the figures in the paper are freely availble at \url{https://github.com/giuliacelora/Active_droplet_chemotaxis.git}.

\appendix
\section{Additional information on the dynamic model and simulations}
\subsection{Non-dimensionalisation of the governing equations}
\label{app:nondim}
Here, we present the steps used to derive the dimensionless modelling equations~\eqref{general governing equations}. Denoting the droplet height, spatial coordinate, and time in dimensional units by $\tilde{h}$, $\tilde{x}$, and $\tilde{t}$, respectively, we perform the following rescaling to obtain Eq.~\eqref{general governing equations}. The spatial variable is rescaled as $\tilde{x} = X x$, where the characteristic length scale is chosen to be the slip length $X=\nu \gg 1$. Time is rescaled according to $\tilde{t} = \tau t$, with characteristic time $\tau = \ell/\bar{U} \ll r^{-1}$, where $\bar{U}$ denotes the characteristic migration speed of the droplet, taken to be the asymptotic speed $u_0$. Physically, $\tau$ corresponds to the time required for the droplet to migrate over the characteristic length $\nu$. We further scale the height $\tilde{h}=Hh$ of the droplet by the characteristic height $H=\nu\tan\theta_e$, where $\tan\theta_e$ is the droplet equilibrium contact angle, which we strongly impose via the dynamic contact angle condition at the boundary~\eqref{general BC}. 
\subsection{Chemoattractant regulation}
\label{app:derivation chemoattractant}
Here, we give more details on the derivation of the functional form of the active forcing~\eqref{eq:active forcing chemotaxis}, starting from a full description of the chemottactrant dynamics. As illustrated in Figure~\ref{fig:model schematic}, we consider a chemoattractant that diffuses and decays at rates $D_c$ and $r_c$, respectively, while being produced at a rate $s_c(x,t)$ by a source that is degraded by the migrating group. For simplicity, we assume the source is instantaneously consumed by the group. Given that we are considering a thin cell group, following~\cite{ford_pattern_2024}, we neglect any dependency of the chemoattractant on the vertical direction, such that $c=c(x,t)$. Under these assumptions, the chemoattractant field $c(x,t)$ evolves in space and time according to  
\begin{subequations}
    \begin{align}
\partial_t c = D_c\partial_{xx}c-r_cc+s_cH(x-x_+(t)), \quad x\in \mathbb{R},\ t>0,\label{eq:chemoattractant}
\end{align}
where $H$ indicates the Heaviside function. We further couple the system with far-field conditions
\begin{align}
    \lim_{x\rightarrow-\infty} \partial_x c(x,t)=0,\quad \lim_{x\rightarrow\infty} c(x,t)=\bar{c}_\infty=\frac{s_c}{r_c}, 
\end{align}\label{sys:chemoattractant}%
\end{subequations}
which corresponds to the equilibration of the reaction terms in the far field~\eqref{eq:chemoattractant}. We further assume that cells respond to the chemoattractant by exerting a force in the direction of increasing chemoattractant concentration. We assume that the force is proportional to the signal measured, here assumed to be the logarithm of the chemoattractant concentration, as in the~\cite{keller_traveling_1971} model of self-generated chemotaxis. Differences in the strength of cell-cell interactions (\emph{i.e.}, force exchange) result in gradients in the active pressure $p_a$ and the generation of an active force $f_a$ 
\begin{align}
p_a=a_0\ln\left(c\right)\quad  \Longrightarrow \quad f_a=\partial_x p_a=\frac{a_0\partial_xc}{c}, \label{eq:activityI}
\end{align}
which is analogous to the functional form for the chemoattractant forcing adopted in the standard Keller-Segel model. However, within the active fluid framework, Eq.~(\ref{eq:activityI}) can be interpreted as the activity profile for an extensile system where the agents are perfectly aligned with the positive gradient direction ($+1$ in the 1D geometry) and the force they exert on each other is proportional to the sensed signal, \emph{i.e.}, the logarithm of the chemoattractant field. 

In general, Eq.~(\ref{eq:activityI}) yields spatio-temporally evolving forces. However, provided that the solution is close enough to its quasi-steady profile, numerical simulations show that the chemoattractant profile quickly equilibrates in the reference frame of the moving droplet, Eq.~(\ref{eq:chemoattractant}). This is because the droplet acceleration $\ddot{x}_+$ is driven by the slow proliferation of the droplet, which is negligible on the timescale of migration (\Cref{sec:main multiple scales}). Moving into the moving frame of the front contact line, Eq.~\eqref{eq:chemoattractant} can be solved explicitly in this quasi-steady limit. We find that the chemoattractant profile decays monotonically towards the rear of the droplet at a rate that is modulated by the speed at which the group migrates
\begin{equation}
    c(x,t)= c_\infty(\dot{x}_+)\,e^{\frac{x-x_+(t)}{\ell_c(\dot{x}_+)}}, \quad x\in(-\infty,x_+(t)]\label{eq:chemoattractant steady}
\end{equation}
where the effective decay rate of the chemoattractant and its maximal concentration within the aggregate are modulated by the speed at which the group advances, \begin{align} 
\ell_c(\dot{x}_+)&=\frac{\dot{x}_+}{2r_c}+\frac{1}{r_c}\sqrt{\left(\frac{\dot{x}_+}{2}\right)^2+r_cD_c}, \\
c_\infty(\dot{x}_+)&=\bar{c}_\infty\frac{\dot{x}_++\sqrt{\dot{x}_+^2+4r_cD_c}}{2\sqrt{\dot{x}_+^2+4r_cD_c}}.
\end{align}
Substituting Eq.~\eqref{eq:chemoattractant steady} into~\eqref{eq:activityI}, we find that, under the quasi-steady assumption, the active pressure $p_a$ generates a spatially-homogeneous force regulated by the length of chemoattractant decay length $\ell_c$
\begin{equation}
    f_a=\frac{a_0}{\ell_c(\dot{x}_+)}. \label{app:eq active forcing 2}
\end{equation}
The expression~\eqref{eq:active forcing chemotaxis} is obtained from~\eqref{app:eq active forcing 2}, normalising $\ell_c$ so that $\ell_c(0)=1$ and introducing the proportionality constant \begin{equation}
\mathcal{A}:=a_0\sqrt{r_c/D_c}\label{app: def A}
\end{equation}
and the parameter \begin{equation}
\eta_c:=1/\sqrt{r_cD_c}.\label{app:def etac}
\end{equation}
\subsection{Numerical simulation}
\label{app:numerics}
We start by investigating the morphology of the migrating droplet under self-generated gradients via numerical simulations of the full problem, coupling the aggregate and chemoattractant dynamics, \emph{i.e.}, Eqs.~(\ref{general governing equations}), \eqref{general BC},  \eqref{sys:chemoattractant} and \eqref{eq:activityI}. 
To solve the thin film governing equation~(\ref{general BC}) numerically, we adopt the same approach as in~\cite{peschka_variational_2018}. This approach leverages an Arbitrary Lagrangian-Eulerian (ALE) reference frame, finite-element methods for the spatial discretisation, and a semi-implicit time-stepping scheme to advance the solution over time. We solve for the chemoattrant profile on a static domain, much larger than the size of the droplet, and use finite differences to discretise Eq.~(\ref{sys:chemoattractant}) in space and the implicit Euler method to advance it in time. An implementation of the approach in \texttt{Julia} is available at~\url{https://github.com/giuliacelora/Active_droplet_chemotaxis.git}. 

For the initial conditions, we set the profile of $h$ to be the corresponding equilibrium static droplet solution of initial volume $v_0$:
\begin{equation}
    h(x,0) = \frac{1}{\ell_0} \left(\frac{3v_0}{2}-x^2\right),
\end{equation}
where $\ell_0$ is the initial length of the droplet, and the initial positions of the front and rear of the droplet are, respectively, $x_\pm(0)=\pm 
\ell_0/2$. We initialise the chemoattractant to be homogeneous $c(x,0)\equiv \bar{c}_\infty=s_c/r_c$. We let the system evolve up to time $t=100$ without any growth, \emph{i.e.,} setting $r=0$ in~\eqref{general governing equations}, to allow the system to relax to the corresponding travelling wave. We call this initial transient the \emph{burn-in} period. After this time, we switch on growth (setting $r$ to the value listed in~\Cref{tab:parameters}) and simulate the system up to a final time past the elongation transition. Results in~\Cref{fig:SG_travelling_homogeneous_example} only illustrate the evolution of the solution after the burn-in period. The parameter values used for the simulation results presented in~\Cref{fig:SG_travelling_homogeneous_example} are given in~\Cref{tab:parameters}.

\begin{table}
    \centering
    \footnotesize
    \begin{tabular}{c p{75mm} c c}
         Parameter & Description & Set 1 & Set 2 \\
         \hline
         $Ca_\kappa$& passive capillary number &$0.2$& $0.02$\\[3pt]
         $\mathcal{A}$& active capillary number (see Eq.~\eqref{app: def A}) &$0.5$& $2$\\[3pt]
         $r$ & droplet proliferation rate & $0.0005$ & $0.0005$\\[3pt]
         $\eta_\theta$& contact line capillary number~\eqref{eq:dynamics_contact angle} &$0.01$&$0.99$\\[3pt]
         $\eta_c$ & ratio between the characteristic timescale and lengthscale of the chemoattractant (see~\eqref{app:def etac}) & $1.0$ & $0.5$\\[3pt]
         \hline
         $a_0$ & activity parameter~\eqref{eq:activityI}&$2$&$200$\\
         $D_c$&chemoattractant diffusion coefficient&$4$&$50$\\[3pt]
         $r_c$&chemoattractant decay rate & $0.25$ &$0.005$\\[3pt]
         $s_c$& rate of chemoattractant production &$0.25$&$0.005$
    \end{tabular}
    \caption{Parameter values used for the simulation results in~\Cref{fig:SG_travelling_homogeneous_example}. Set 1 corresponds to the example ``smooth morphological transition'', while Set 2 corresponds to the example ``sharp morphological transition''. Parameters above the horizontal line are the ones imposed for solving~\eqref{general governing equations}; the parameters below the line are used to calculate $\mathcal{A}$ and $\mathcal{\eta}_c$ via Eqs.~\eqref{app: def A}-\eqref{app:def etac}. Based on the listed values of the model parameters, the corresponding non-dimensional input parameters for the elongated-droplet theory (see~\Cref{app:self-generated chemotaxis_conditions}) are: (Set 1) $\gamma_\theta\approx0.99$ and $b=0.45$; (Set 2) $\gamma_\theta\approx 0.03$ and $b=0.71$.} 
    \label{tab:parameters}
\end{table}
\subsection{Numerical continuation of TW solutions}
\label{app:numerical continuation}
We describe here the numerical procedure used to obtain the bifurcation diagrams shown in Figures~\ref{fig bifurcation constant forcing panel A}–\ref{fig bifurcation constant forcing panel B}. Equation~\eqref{eq:general_TW_problem} is solved for a prescribed value of $\ell$ using a collocation method, and numerical continuation is then employed to track how the solution evolves as a function of $\ell$. The implementation is carried out in \texttt{Julia}, making use of the \texttt{BifurcationKit.jl} package for continuation. The collocation method is based on a finite-difference discretisation of~\eqref{general governing equations} on a uniform mesh. The code is available at~\url{https://github.com/giuliacelora/Active_droplet_chemotaxis.git}. 

\section{Method of multiple scales}
\label{app_sec:multiple_scales}
Here, we present the derivation of the system~\eqref{sys:h_multiple_scales} from~\eqref{general governing equations}-\eqref{general BC} in the limit of slowly proliferating cells, namely $r\ll1$. In this regime, the timescales of migration $t_u=\bar{u}^{-1}\sim\mathcal{O}(1)$ as $r\to 0$ and growth $t_{p}=r^{-1}\gg 1$ separate, allowing us to derive an effective model that describes the migration dynamics as controlled by the slowly evolving droplet volume. We introduce the ``slow" timescale $T$ via $T=rt$, and refer to the original timescale $t\sim\mathcal{O}(1)$ as the ``fast" timescale. Using the method of multiple scales~\citep{bender_multiple-scale_1999}, we treat the two as independent and transform the time derivative as
\begin{equation}
    \partial_t \mapsto \partial_t + r\partial_T, \label{eq:time_map_multiple_scales}
\end{equation}
so that Eqs.~\eqref{general governing equations}-\eqref{general BC} transform to
\begin{subequations}\label{sys_multiple_scales}
    \begin{align}
        \partial_t h +r\partial_Th&= -\partial_x\left(h\bar{u}(h)\right)+rh, \quad x\in[x_-,x_+],\\[5pt]
          \left.h\right|_{x=x_\pm}&=0,\quad  \left.h(\partial_tx_\pm+r\partial_Tx_{\pm}-\bar{u}(h))\right|_{x=x_\pm}=0,\\ 
          \eta_\theta (\partial_tx_\pm+r\partial_Tx_{\pm})&=\pm \left[\left(\left.\partial_x h\right|_{x=x_\pm}\right)^2-1\right], 
    \end{align}
    \end{subequations}
    with appropriate initial conditions for both fast and slow timescales. 
We expand the droplet height profile as an asymptotic series in powers of $r$ and functions of both the fast and slow timescales
\begin{equation}
h=h_0(x,t,T)+rh_1(x,t,T)+\mathcal{O}(r^2),\label{app:multiple_scales_expansion}
\end{equation}
and recover at leading-order the purely transport equation
\begin{subequations}\label{sys_multiple_scales:leading_order}
\begin{align}
    \partial_th_0+\partial_x(h_0\bar{u}_0)=0,\quad x\in[x_-,x_+],\label{app_multiple_scales:leading_order_hevol}\\
    \left.h_0\right|_{x=x_\pm}=0, \quad \left.h_0(\partial_tx_\pm-\bar{u}(h_0))\right|_{x=x_\pm}=0, \label{app_multiple_scales:leading_order_noflux}\\
    \eta_\theta \partial_tx_\pm=\pm \left[\left(\left.\partial_x h_0\right|_{x=x_\pm}\right)^2-1\right].
\end{align}
     Integrating Eq.~\eqref{app_multiple_scales:leading_order_hevol} and imposing the no-flux conditions~\eqref{app_multiple_scales:leading_order_noflux}, we find that the leading-order approximation of the droplet volume is constant in the fast timescale and only evolves in the slow timescale
     \begin{align}
         \int_{x_-}^{x_+}h_0(x,t,T)dx=v_0(T).\label{app_multiple_scales:leading_order_mass_constrain}
     \end{align}
     \end{subequations}
From the leading-order problem~\eqref{sys_multiple_scales:leading_order}, we see that the evolution of the leading-order approximation of the droplet profile has no explicit dependence on the slow timescale $T$, which enters as an effective control parameter via the mass constraint~\eqref{app_multiple_scales:leading_order_mass_constrain}. To determine $h_0=h_0(t,x;T)$, we must therefore derive the equation governing the evolution of the volume $v_0(T)$ in the slow timescale. To obtain this, we must proceed to the next asymptotic order of Eq.~\eqref{sys_multiple_scales} (\emph{i.e.}, at $\mathcal{O}(r)$), and derive the solvability condition. At the next order, we have
\begin{subequations}
   \begin{align}
        \partial_th_1+\partial_Th_0&=-\partial_x\left(h_1\bar{u}(h_0)+h_0\bar{u}'(h_0)[h_1]\right)+h_0,\quad x\in[x_-,x_+],\label{app_multiple_scales:first_correction_order_hevol}\\[5pt]
         \left. h_{1}\right|_{x_\pm}&=0,\quad   \left[h_1(\partial_T x_\pm-\bar{u}(h_0))+h_0(\partial_tx_\pm-\bar{u}'(h_0)[h_1])\right]_{x=x_\pm}=0,\label{app_multiple_scales:first_correction_order_noflux} \\
             \eta_\theta \partial_Tx_\pm&=\pm 2\left.\partial_x h_0\partial_x h_1\right|_{x=x_\pm},
    \end{align}
    \end{subequations}
    where $\bar{u}'(h_0)[\cdot]$ is the linear differential operator
    \begin{equation}
    \bar{u}'(h_0)[h_1]=h_0\partial_{xxx}h_1+h_1\partial_{xxx}h_0+h_1f_a(\partial_t x_+)+\partial_Tx_+ h_0 f_a'(\partial_t x_+).
    \end{equation}
    We integrate~\eqref{app_multiple_scales:first_correction_order_hevol} over the spatial variable $x$ and imposing the no-flux condition~\eqref{app_multiple_scales:first_correction_order_noflux}, to obtain
    \begin{equation}
    \partial_tv_1=v_0-\partial_Tv_0,\label{app_multiple_scales:first_correction_mass_constrain}
    \end{equation}
    where $v_1(t,T)=\int_{x_-}^{x_+}h_1(x,t,T)dx$. In the standard method of multiple scales, $v_1$ is required to be periodic in the ``fast" timescale (here, this implies that $v_1$ has to be constant), which leads to the solvability condition:
    \begin{equation}
        \partial_Tv_0=v_0 \Rightarrow v_0=v(0)\exp(T).\label{app_multiple_scales:first_correction_solvability}
    \end{equation}
    Combining~\eqref{sys_multiple_scales:leading_order} with the slow-time evolution~\eqref{app_multiple_scales:first_correction_solvability}, we obtain the system~\eqref{sys:h_multiple_scales}-\eqref{eq:volume constaint_slow_variable}.
\section{Droplet volume}\label{app:mass calculation}
Here, we briefly derive~\eqref{def volume dimensional}, which relates the droplet volume to its length and migration speed. First, we can compute the volume of the droplet in terms of its other geometrical properties by writing~\eqref{eq:H_TW1} in the following conservative form
\[\tilde{h}\tilde{h}'''=\left(\tilde{h}\tilde{h}''-\frac{(\tilde{h}')^2}{2}\right)'=Ca_\kappa u-f_a(u)\tilde{h}.\]
 Integrating the above expression, we find
 \begin{equation}
 \begin{aligned}f_a(u)\int_{-\ell/2}^{\ell/2}\tilde{h}(\phi) d\phi-uCa_\kappa \ell&=\left[-\tilde{h}\tilde{h}''+\frac{(\tilde{h'})^2}{2}\right]^{\ell/2}_{-\ell/2}\\&=\frac{(h'(\frac{\ell}{2}))^2-(h'(-\frac{\ell}{2}))^2}{2}=\eta_\theta u,
 \end{aligned}
 \end{equation}
 from which the expression~\eqref{def volume dimensional} for the volume $v=\int_{-\ell/2}^{\ell/2}\tilde{h}(\phi) d\phi$ naturally follows.
\section{Small-droplet TW solutions}
\label{app: small droplet limit}
Here, we detail the procedure to derive the approximate small-droplet solution for the non-linear eigenvalue problem~(\ref{eq:general_TW_problem}), which has been used to produce Figure~\ref{fig bifurcation constant forcing}. To capture this regime, we introduce the small parameter $\varepsilon=\ell^2$ and the following scalings for the other unknowns 
\begin{equation}
    u=\frac{U}{Ca_\kappa}\sqrt{\varepsilon}, \quad \phi=Z\sqrt{\varepsilon}, \quad h=H\sqrt{\varepsilon}, \quad v=\varepsilon V.
\end{equation}
We seek asymptotic solutions of the form $H\sim H_0+\varepsilon H_1$, $U=U_0+\varepsilon U_1$, and $V=V_0+\varepsilon V_1$. The leading-order version of Eq.~(\ref{eq:general_TW_problem}) is the corresponding equilibrium formulation for a stationary droplet shaped by surface tension:
\begin{subequations}
    \begin{align}
        H'''_0&=0,\quad Z\in\left[-\frac{1}{2},\frac{1}{2}\right]\\
        H_0\left(\pm\frac{1}{2}\right)&=0,\quad H_0'\left(\pm\frac{1}{2}\right)=\mp1,
        \end{align}\label{eq:leading-order}
\end{subequations}
which has the explicit solution
\begin{equation}
    H_0(Z)=\left(Z+\frac{1}{2}\right)\left(\frac{1}{2}-Z\right).
\end{equation}
Having an explicit solution for $H_0$ we can directly calculate $V_0$ as
\begin{equation}
V_0=\frac{1}{6},
\end{equation}
and use this to estimate the leading-order contribution to the migration velocity using~\eqref{def volume dimensional},
\begin{equation}
    U_0=\frac{1}{6}\frac{\mathcal{F}_a(U_0)}{1+\mathcal{N}_\theta},\label{eq:velocity_asymptotics}
\end{equation}
where we define $\mathcal{N}_\theta:=\varepsilon^{-1/2}Ca_\kappa\eta_\theta$ and $\mathcal{F}_a(U)=f_a(UCa_\kappa\sqrt{\varepsilon}
)$. Eq.~\eqref{eq:velocity_asymptotics} implicitly defines $U_0$, whose value is unique since $f_a$ is a positive and monotonically decreasing function of $U_0$ by assumption~\eqref{eq:active forcing chemotaxis}. 
The contribution of the active force on the aggregate dynamics enters into the problem for the first-order approximation $H_1$
\begin{subequations}
    \begin{align}
       H'''_1&= \frac{U_0}{ H_0} - f_{a}(U_0),\quad Z\in\left[-\frac{1}{2},\frac{1}{2}\right]\label{eq:H_second_order}\\
        H_1\left(\pm\frac{1}{2}\right)&=0,\quad H_1'\left(\pm\frac{1}{2}\right)=-\frac{1}{2}\mathcal{N}_\theta U_0,\label{eq:bc_second order}
    \end{align}\label{eq:second order}%
\end{subequations}
where $U_0$ is now known via~\eqref{eq:velocity_asymptotics}. Since Eq.~\eqref{eq:H_second_order} is a third-order ODE, the four boundary conditions~\eqref{eq:bc_second order} may appear to overdetermine the system. However, only three of the four boundary conditions are independent once the leading-order velocity $U_0$ is set to satisfy~\eqref{eq:velocity_asymptotics}. Therefore, we can impose any three of the four boundary conditions, and the latter is naturally satisfied. Here, without loss of generality, we impose the two boundary conditions on $H_1'$ and the Dirichlet condition at $Z=-1/2$ and explicitly integrate~(\ref{eq:H_second_order}) to find the $\mathcal{O}(\varepsilon)$ solution
\begin{subequations}
\begin{align}
    H_1(Z)= &\frac{U_0}{2}\left[\psi(Z)-\frac{1+2Z}{2}\right]-\frac{\mathcal{N}_\theta U_0\left(2Z+1\right)}{4}+\frac{f_a(U_0)(1+2Z)^2(1-Z)}{24},\label{eq:H_first_order_correction}
    \end{align}
where we define 
    \begin{align}
        \psi(Z)&=\frac{\left(1+2Z\right)^2}{4}\log\left(\frac{1+2Z}{2}\right) - \frac{(1-2Z)^2}{4}\log\left(\frac{1-2Z}{2}\right).
    \end{align}
\end{subequations}
In Eq.~(\ref{eq:H_first_order_correction}), the first term arises from the deformation of the droplet shape due to the fact that it is moving, the second term comes from the contact line dynamics, and the last two terms are associated with the active forcing. We emphasise that~\eqref{eq:H_first_order_correction} satisfies the Dirichlet condition at $Z=1/2$ provided $U_0$ satisfies~\eqref{eq:velocity_asymptotics}. Finally, we integrate Eq.~(\ref{eq:H_first_order_correction}) over $Z\in(-1/2,1/2)$ to find that the first-order correction to the volume $V_1$, and, therefore, the velocity $U_1$ both vanish.

Writing the expression in terms of the original model variables, and considering the specific form of the forcing~\eqref{eq:active forcing chemotaxis}, we find 
\begin{equation}
\begin{aligned}
    v&=\frac{\ell^2}{6}(1+\mathcal{O}(\ell^4)), \quad u=\frac{\mathcal{A}\ell^2(1+\mathcal{O}(\ell^4))}{Ca_\kappa\sqrt{\ell+\eta_\theta}\sqrt{36\ell+6\mathcal{A}\eta_c\ell^2+36\eta_\theta}}. \label{eq:mass_asymptotics}
    \end{aligned}
\end{equation}
Eq.~(\ref{eq:mass_asymptotics}) highlights the relationship between the volume of the droplet, its length, the active force at the substrate, and active stresses in the bulk. Figure~\ref{fig bifurcation constant forcing} shows a comparison between~\eqref{eq:mass_asymptotics} and the numerically estimated travelling wave speed and volume for different values of the parameters. As expected, the asymptotic theory agrees with the behaviour of the $u(\ell)$ and $v(\ell)$ for sufficiently small values of the droplet length (and volume), but fails to capture the saturation of the travelling speed, which is due to changes in the morphology of the aggregate shape.

\section{Additional results for elongated TW analysis}
\label{app:computation BL}
\begin{figure}
    \centering
   \centering
           \begin{subfigure}{0.00\textwidth}
    \captionlistentry{}
    \label{fig: app velocity}
    \end{subfigure}
               \begin{subfigure}{0.00\textwidth}
    \captionlistentry{}
    \label{fig: app volume}
    \end{subfigure}
    \includegraphics[width=\linewidth]{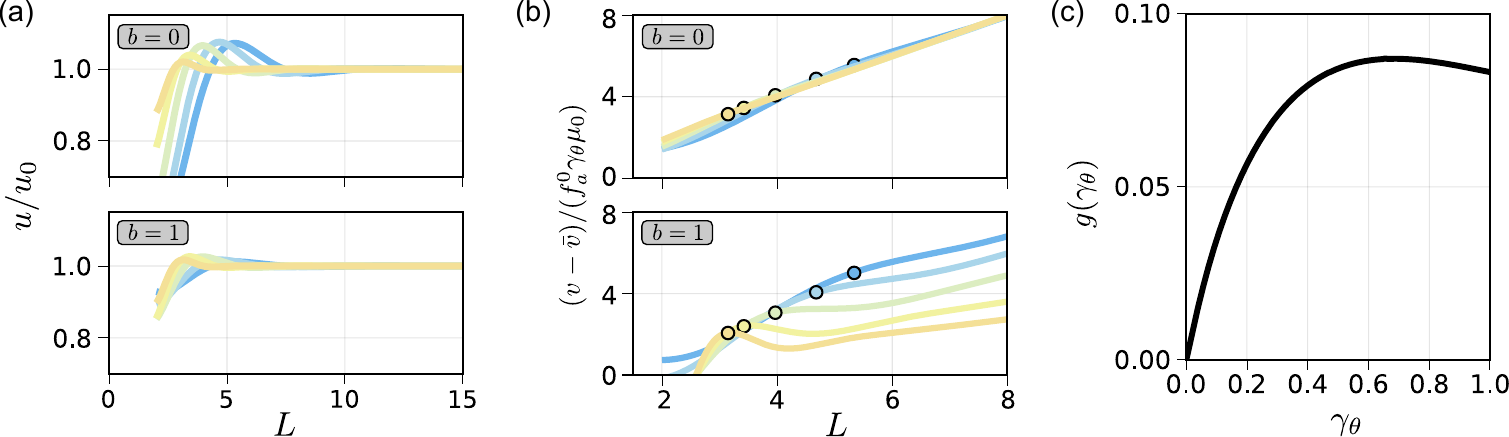}
    \begin{subfigure}{0.00\textwidth}
    \captionlistentry{}
    \label{fig: app g}
    \end{subfigure}
    \vspace{-5mm}
    \caption{(a) Relation between the non-dimensional droplet length ($L$) and the droplet migration speed $u$ (up to first-order corrections~\eqref{eq leading-order function of L}) for different values of the parameter $b$: (top) no coupling between chemoattractant and migration speed ($b=0$); (bottom) strong coupling between chemoattractant and migration speed ($b=1$). The colour indicates different values of $\gamma_\theta$ (or equivalently $\mathcal{N}_\theta$ as in~\Cref{fig:regionID}). (b) Same as (a) but for the droplet volume $v$~\eqref{eq:volume up to first-order correction}. The constant $\bar{v}=1-\gamma_\theta^{4/3}$ indicates the offset in the leading-order approximation of the length-volume relation~\eqref{eq:volume leading-order}. (c) Plot of the function $g(\gamma_\theta)$ defined by~\eqref{app:eq g}. }
    \label{fig:placeholder}
\end{figure}

Here, we collect additional results needed to complete the derivation of the elongated droplet solutions presented in~\Cref{sec analysis:summary}. Figures~\ref{fig: app velocity}-\ref{fig: app volume} illustrate the mapping between the rescaled droplet length $L$ and its speed and volume as predicted by~\eqref{eq leading-order function of L} and~\eqref{eq:volume up to first-order correction}, respectively.

\subsection{Numerical approximation of the droplet profile in the boundary layers}
\label{app:numerics for asymptotics}
Given that, in boundary-layer problems, the information typically propagates from the far field, we employ a shooting method to resolve the droplet profile. Specifically, we initialise the integration at a large but finite value of $\sigma$, using the WKBJ analysis to accurately characterise the far-field behaviour, and then use a standard time-stepping scheme for ODEs in \texttt{DifferentialEquations.jl} to obtain a numerical approximation of the height profile. The code is available at~\url{https://github.com/giuliacelora/Active_droplet_chemotaxis.git}.
\subsubsection{Leading-order solution in the rear boundary layer}
\label{app: Leading-order solution in the rear boundary layer}
We approximate the solution to~(\ref{problem constant inner rear leading-order}), solving Eq.~(\ref{eq:Y0 rear}) with a different set of boundary conditions:
\begin{subequations}
    \begin{align}
        \BLI{H}_0(\sigma^*)=1-e^{-\sigma^*}, \quad \BLI{H}_0'(\sigma^*)=-e^{-\sigma^*}, \quad \BLI{H}_0''(\sigma^*)=e^{-\sigma^*}
    \end{align}\label{sys:shooting Y0 rear}
\end{subequations}
where $\sigma^*=10$ is chosen arbitrarily to be sufficiently large. We integrate Eq.~(\ref{eq:Y0 rear}) backwards ($\sigma<\sigma^*$) until we reach the point $\sigma_0$ at which $\BLI{H}_0(\sigma_0)\approx0$, using an absolute tolerance of $10^{-15}$. We then translate the obtained solution to satisfy the boundary conditions~\eqref{far field outer rear} using the transformation $\sigma \hookrightarrow\sigma-\sigma_0$ and obtain the constant $\mu_0$ as
\begin{equation}
    A_0=-e^{\sigma_0}, \quad \mu_0=\frac{1}{\BLI{H}'(0)^{2/3}},
\end{equation}
ensuring that $\sigma^*\gg\sigma_0$.
\subsubsection{Leading-order solution in the front boundary layer}
\label{app: Leading-order solution in the front boundary layer}

To resolve the front boundary layer, we adapt the strategy outlined in~\Cref{app: Leading-order solution in the rear boundary layer} to account for the additional complication that the phase shift $\BLII{\psi}_0$ is a priori unknown. Specifically, $\BLII{\psi}_0$ is selected by the behaviour in the rear boundary layer via the dependency of the boundary condition~\eqref{front leading-order BC} on the velocity $U_0$. Rather than approximating the mapping $\tilde{\psi}_0(U_0)$, it is numerically more straightforward to estimate its inverse, \emph{i.e.,} $U_0(\BLII{\psi}_0):[-\pi,\pi)\rightarrow (0,\infty)$. To do this, we discretise the interval $[-\pi,\pi)$ into 7000 equally spaced points and, for each value of $\BLII{\psi}^{n}_0$, we integrate Eq.~\eqref{eq:Y0 leading front} forward in space with boundary conditions
\begin{equation}
\begin{aligned}
        \BLII{H}_0(\sigma^*;n)&=1+e^{-\sigma^*/2}\sin\left(-\frac{\sqrt{3}}{2}\sigma^*+\BLII{\psi}^{n}_0\right),\\ \BLII{H}_0'(\sigma^*;n)&=e^{-\sigma^*/2}\sin\left(-\frac{\sqrt{3}}{2}\sigma^*+\BLII{\psi}^n_0+\frac{\pi}{3}\right), \\\BLII{H}_0''(\sigma^*;n)&=e^{-\sigma^*/2}\sin\left(-\frac{\sqrt{3}}{2}\sigma^*+\BLII{\psi}^n_0+\frac{2\pi}{3}\right),
        \end{aligned}
\end{equation}
where $\sigma^*\gg1$.
We integrate Eq.~(\ref{eq:Y0 leading front}) forward ($\sigma>-\sigma^*$) until we reach the point $\BLII{\sigma}^n_0$ at which $\BLII{H}_0(\BLII{\sigma}^n_0;i)\approx0$, using a tolerance of $10^{-15}$.  We then translate the solution to satisfy the boundary conditions~\eqref{BCI:leading} using the transformation $\sigma \hookrightarrow\sigma-\BLII{\sigma}^n_0$. We directly obtain the constant $\BLII{\sigma}^n_0$ in Eq.~\eqref{far field outer front} and use the condition of $\BLII{H}'_0$ in Eq.~(\ref{BCI:leading}) to derive the corresponding value of the velocity $U_0$
\begin{equation}
    U^n_0=U_0(\BLII{\psi}^n_0)=\mu_0\left(\frac{2}{1+(\BLII{H}'_0(0))^2\mu^{4/3}_0}\right)^{3/4}.\label{app:velocity phase relationship rear}
\end{equation}
We can then use the expression~\eqref{eq constant: U_0} to obtain an implicit map between $\BLII{\psi}_0$ and $\mathcal{N}_\theta$. The latter is illustrated in~\Cref{fig:regionIB}. We note that $\BLII{\psi}_0$ takes value in a subset of the period $[-\pi,\pi)$ and asymptotes towards a finite angle $\BLII{\psi}_0^{cr}$ as $\mathcal{N}_\theta\to\infty$.
\subsubsection{First-order correction in the rear boundary layer}
\label{app:solve first order correction rear}
Here, we detail the approach used to determine the first-order correction in the rear boundary layer. We exploit the linearity of Eqs.~(\ref{eq:Y1 rear}) and write $\BLI{H}_1$ as
\begin{equation}
\BLI{H}_1=B^{\dagger}e^{\sigma/2}\Imag\left[\exp\left[i\left(\psi^\dagger_0-\frac{\sqrt{3}}{2\varepsilon U_0^{1/3}}\right)\right]\BLI{Y}_B\right]+\BLI{A}_1\BLI{Y}_A,\label{app: eq expansion H1}
\end{equation}
where the complex function $\BLI{Y}_B$ satisfies
\begin{equation}
\BLI{Y}_B'''+\frac{3}{2}\BLI{Y}_B''+\frac{3}{4}\BLI{Y}_B'+\left(\frac{1}{8}+\frac{1}{\BLI{H}_0^2}\right)\BLI{Y}_B=0, \quad \lim_{\sigma\rightarrow\infty} \BLI{Y}_B\sim \exp\left[i\frac{\sqrt{3}}{2}\sigma\right],\label{eq:YBI}
\end{equation}
and the real function $\BLI{Y}_A$ satisfies
\begin{equation}
    \BLI{Y}'''_A+\frac{\BLI{Y}_A}{\BLI{H}_0^2}=0, \quad \lim_{\sigma\rightarrow \infty}\BLI{Y}_A\sim e^{-\sigma}.\label{eq:YAI}
\end{equation}
Since the exponentially large far-field matching is numerically challenging to deal with, in writing~\eqref{app: eq expansion H1} we have factored out the exponential term $e^{\sigma/2}$ for numerical stability. We solve the problems~\eqref{eq:YBI}-\eqref{eq:YAI} using a shooting approach from $\sigma=\sigma^*\gg1$ coupled to the leading-order problem for $\BLI{H}_0$ (see~\eqref{eq:Y0 rear}). Recalling that $B^{\dagger}$ and $\psi^\dagger$ are known and obtained via matching the solutions in the outer and front boundary layer~\eqref{matching outer front BL}, imposing the boundary conditions~\eqref{BC first correction rear} at $\sigma=0$  allow us to easily compute the value of the unknown constants $\BLI{A}_1$ and $U_1$
\begin{subequations}
    \begin{align}
        \BLI{A}_1&= - \frac{B^\dagger|\BLI{z}|\sin\left(\psi^\dagger+\arg \BLI{z}\right)}{\BLI{Y}_A(0)},\\
    U_1&= -\frac{B^\dagger|\BLI{s}|6\mu^{2/3}_0U_0}{\dfrac{4-\mathcal{N}_\theta U_0}{1-\mathcal{N}_\theta U_0}-2U_0\mathcal{F}'_a(U_0)}\sin\left(\psi^\dagger+\arg\BLI{s}\right),\label{app:U1}
    \end{align}
\end{subequations}
where we numerically determine the complex constants
\begin{equation}
\BLI{z}=\BLI{Y}_B(0)\approx 29.165 + 118.516i
\end{equation}
and 
\begin{equation}
\BLI{s}=\BLI{z}\left(\frac{1}{2}-\dfrac{\BLI{Y}_A'(0)}{\BLI{Y}_A(0)}+\dfrac{\BLI{Y}_B'(0)}{\BLI{Y}_B(0)}\right)\approx 0.824 + 0.106 i.
\end{equation}
 The expression~\eqref{eq:U1} is obtained from~\eqref{app:U1} by simplying the denominator using~\eqref{eq constant: U_0}.
\subsubsection{First-order correction in the front boundary layer}
The correction in the front boundary layer $\BLII{H}_1$ is determined by
\begin{subequations}\label{eq:Y correction front}
    \begin{align}
        \BLII{H}'''_1+\frac{\BLII{H}_1}{\BLII{H}_0^2}&=0,    \label{eq:ODE first order correction front} \\
        \BLII{H}_1(0)=0, \quad \BLII{H}_1'(0)&=\left(\frac{4+\mathcal{N}_\theta U_0}{1+\mathcal{N}_\theta U_0}-2U_0\mathcal{F}'_a(U_0)\right)\frac{U_1}{6U_0\mu^{2/3}_0}\sqrt{\frac{2\mu^{4/3}_0}{U^{4/3}_0}- 1},\label{eq BC first order correction front}\\
        \lim_{\sigma\rightarrow -\infty}\BLII{H}_1&=0, \label{eq:far field first correction front}
    \end{align}
\end{subequations}
where the far field condition~\eqref{eq:far field first correction front} is obtained by matching with the correction at $\mathcal{O}(1/(2\varepsilon U_0^{1/3}))$ in the outer solution~\eqref{far field outer front}. Here, the condition~\eqref{eq:far field first correction front} imposes a single independent constraint, which can be shown via a WKBJ analysis of the far-field behaviour of~\eqref{eq:ODE first order correction front}. Since~\eqref{eq:ODE first order correction front} is a third-order ODE, $U_0$ and $U_1$ are known, and the boundary conditions~\eqref{eq BC first order correction front}-\eqref{eq:far field first correction front} impose three independent constraints, we conclude that the problem is well-posed. 

It is possible to solve~\eqref{eq:Y correction front} using a similar approach to the one described in~\Cref{app:solve first order correction rear}, which exploits the linearity of~\eqref{eq:ODE first order correction front} to simplify the shooting procedure. In particular, we write the solution as
\[\BLII{H}_1=\BLII{B}_1\Imag\left[e^{i\BLII{\psi}_1}\BLII{Y}_B\right],\]
where $\BLII{B}_1$ and $\BLII{\psi}_1\in[-\pi,\pi)$ are unknown constants that depend on $U_0$ and $U_1$, while the complex function $\BLII{Y}_B$ satisfies the third-order ODE
\begin{equation}
    \BLII{Y}_B'''+\frac{\BLII{Y}_B}{\BLII{H}_0^2}=0, \quad \lim_{\sigma\rightarrow -\infty}\BLII{Y}_B\sim \exp\left[\frac{1+i\sqrt{3}}{2}\sigma\right].\label{eq: shooting first order correction front}
\end{equation}
We solve the problem by integrating Eqs.~\eqref{eq:Y0 leading front} and~\eqref{eq: shooting first order correction front} from a large, yet finite, $-\sigma^*\gg1$ for different values of the leading-order approximation of the phase $\BLII{\psi}_0$. Imposing the boundary conditions~\eqref{eq BC first order correction front} at $\sigma=0$ then determines the unknowns $\BLII{B}_1$ and $\BLII{\psi}_1$ in terms of known quantities
\begin{equation}
\BLII{B}_1=  \left(\frac{4+\mathcal{N}_\theta U_0}{1+\mathcal{N}_\theta U_0}-2U_0\mathcal{F}'_a(U_0)\right) \frac{U_1\sqrt{\frac{2\mu^{4/3}_0}{U^{4/3}_0}- 1}}{6  |\BLII{z}|U_0\mu_0^{2/3}\Imag{\,\BLII{s}}},\quad \BLII{\psi}_1=-\arg\BLII{s}.\label{eq: first order correction constants BL front}
\end{equation}
In Eq.~\eqref{eq: first order correction constants BL front}, we have introduced the two complex numbers 
\begin{equation}
\BLII{s}=\frac{\BLII{Y}_B'(0)}{\BLII{Y}_B(0)} \text{ and } \BLII{z}=\BLII{Y}_B(0),
\end{equation}
whose value depends on the phase shift $\BLII{\psi}_0$, which is itself determined by the contact line dissipation parameter $\mathcal{N}_\theta$, as illustrated in~\Cref{fig:regionIC}.
\subsection{Limit of validity of the theory}
\label{computation g}
In this section, we detail the derivation of the expression~\eqref{eq leading-order function of L} and the limits of its validity. Combining Eqs.~\eqref{eq:scaling},~\eqref{eq:general exponential expansion} and~\eqref{eq:U1}, we find that
\begin{equation}
    \frac{u(L)}{u_0}\sim 1+\frac{6 \mu_0^{2/3}B_0^{\dagger}|\BLI{s}|\exp\left[-\dfrac{L}{2 U_0^{1/3}}\right]}{1+\dfrac{3\mu_0^{4/3}}{U^{4/3}_0}+2b}\sin\left(\dfrac{\sqrt{3}L}{2 U_0}-\psi^\dagger-\arg\BLI{s}\right).\label{app: expression u up to first correction}
\end{equation}
We obtain the expression for $L_{cr}$ by considering the extremal point of~\eqref{app: expression u up to first correction}
\begin{equation}
    L_{cr}=\frac{2 U_0^{1/3}}{\sqrt{3}}\left(\psi^\dagger+\arg\BLI{s}+\frac{\pi}{3}\right).\label{app:eq critical length}
\end{equation}

Substituting~\eqref{app:eq critical length} and~\eqref{eq:Ntheta_gamma_theta} into~\eqref{eq leading-order function of L}, and comparing the obtained expression with~\eqref{app: expression u up to first correction}, we find that the function $g$ is defined as
\begin{equation}
    g(\gamma_\theta)=\frac{6\mu_0^{2/3}\gamma_\theta^{4/3}|\BLI{s}|B^\dagger(\gamma_\theta)}{\gamma_\theta^{4/3}+3}\exp\left[-\frac{\psi^{\dagger}(\gamma_\theta)+\arg\BLI{s}+\pi/3}{\sqrt{3}}\right].\label{app:eq g}
\end{equation}
In writing~\eqref{app:eq g}, we have made explicit the dependence of $B^\dagger$ and $\psi^\dagger$ on the strength of the contact line dissipation via the parameter $\gamma_\theta$, which is related to $\mathcal{N}_\theta$ via the bijective map~\eqref{eq:Ntheta_gamma_theta}. The numerically estimated profile of $g$ is shown in~\Cref{fig: app g}. From the plot, we notice that $g<1$, indicating that the first-order correction to the droplet speed is subdominant whenever $L\gtrsim L_{cr}$ (see~\eqref{eq leading-order function of L}). This supports the observation that the elongated droplet limit captures the droplet morphological transition from a compact to an elongated profile.

\bibliographystyle{jfm}
\bibliography{jfm-instructions}

\end{document}